\documentclass[aps,twocolumn,nofootinbib,groupedaddress,longbibliography]{revtex4-1}

\usepackage{epic}
\usepackage{graphicx}
\usepackage[T1]{fontenc}
\usepackage{stackrel}
\usepackage{float}
\usepackage{subfig}
\usepackage[utf8x]{inputenc}
\usepackage{amssymb,amsmath,amsfonts}
 \newcommand{\be}{\begin{equation}}
 \newcommand{\ee}{\end{equation}}
 \newcommand{\bse}{\begin{subequations}}
 \newcommand{\ese}{\end{subequations}}
\newcommand{\ba}{\begin{eqnarray}}
\newcommand{\ea}{\end{eqnarray}}
\renewcommand{\(}{\left(}
 \renewcommand{\)}{\right)}

\usepackage{color}
\usepackage[normalem]{ulem}  
\newcommand{\dt}{\boldsymbol{\cdot}}
\newcommand{\Q}{\mathcal{Q}}
\newcommand{\q}{\mathsf{Q}}

\newcommand{\bQ}{\boldsymbol{\mathcal{Q}}}
\newcommand{\bq}{\boldsymbol{\mathsf{Q}}}
\newcommand{\me}{\mathcal{E}}

\newcommand{\si}{\sigma}
\newcommand{\la}{\lambda}

\begin{document}

\title{$SL(2,R)$ lattices as information processors}
\author{Tanay Kibe}
\email{tanaykibe2.71@gmail.com}
\affiliation{Department of Physics, Indian Institute of Technology Madras,\\ Chennai 600036, India}
\author{Ayan Mukhopadhyay} 
\email{ayan@iitm.ac.in}
\affiliation{Department of Physics, Indian Institute of Technology Madras,\\ Chennai 600036, India}
\author{Alexander Soloviev}
\email{alexander.soloviev@stonybrook.edu}
\affiliation{Department of Physics and Astronomy, Stony Brook University, \\ Stony Brook, New York 11794, USA}
\author{Hareram Swain}
\email{dhareram1993@gmail.com}
\affiliation{Department of Physics, Indian Institute of Technology Madras,\\ Chennai 600036, India}

\begin{abstract}
Black holes past their Page times should act as efficient scramblers and information mirrors. The information of the infalling bits are rapidly encoded by the old black hole in the Hawking quanta, but it should take time that is exponential in the Page time entropy to decode the interior. Motivated by the features of fragmentation instability of near-extremal black holes, we construct a simple phenomenological model of the black hole as a lattice of interacting nearly $AdS_2$ throats 
with gravitational hair charges propagating over the lattice. We study the microstate solutions and their response to shocks. The energy of the shocks are almost wholly absorbed by the total ADM mass of the $AdS_2$ throats, but the information of their locations and time-ordering come out in the hair oscillations, which decouple from the final microstate to which the full system quickly relaxes. We discuss the Hayden-Preskill protocol of decoding infalling information. We also construct generalizations of our model involving a lattice  of $AdS_2$ throats networked via wormholes and their analogues in the form of tensor networks of SYK spin-states.
\end{abstract}


\maketitle


\section{Introduction}\label{Sec:Intro}

Black holes past their Page times are very intriguing information processors. After the Page time, the entanglement entropy of the \textit{Hawking radiation} \cite{Hawking:1975aa,Hawking:1976} should decrease, implying that the information of the formation of the black hole should gradually be revealed  \cite{Page:1993wv,Page:2013dx}. Assuming that the state of the old black hole is typical and that the evaporation process is unitary, one is led to the hypothesized \textit{Page curve} \cite{Page:1993wv,Page:2013dx} for the time-dependence of the entanglement entropy of the Hawking radiation, which needs to be reconciled with Hawking's original computation \cite{Hawking:1975aa}. However, there are other information paradoxes, discussed by Mathur \cite{Mathur:2009hf}, Braunstein, Pirandola and  Zyczkowski \cite{Braunstein:2009my} and Almheiri, Marolf, Polchinski and Sully \cite{Almheiri:2012rt}, which show that all the postulates of \textit{black hole complementarity} \cite{Susskind:1993if} cannot hold simultaneously (see \cite{Harlow:2014yka} for a review). 
One convincing way to avoid these without requiring any breakdown of the local equivalence principle is to argue following Harlow and Hayden that the Hawking radiation post Page time has sufficient quantum complexity \cite{Harlow:2013tf}. The information of the black hole interior embedded in it must take timescales that are \textit{exponential} in the entropy (at the Page time) to be processed even if the most efficient decoders are employed. The implication is that the black hole interior information processing time has to be larger than the evaporation time of the black hole which scales \textit{polynomially} with the entropy at Page time. This Harlow-Hayden argument has been  recently sharpened with reasonable assumptions such as pseudorandomness in the post Page time Hawking radiation \cite{Kim:2020cds}. 

Such models of information processing 
lead to a tantalizing conclusion  that the information of  the bits infalling into the black hole after the Page time should come out in the Hawking radiation rather quickly \cite{Hayden:2007cs}. The Hayden-Preskill timescale, after which this information of an infalling qubit can be decoded from the full radiation Hilbert space including both the early and the late radiation, should scale as $r_s \, \log r_s$ with $r_s$ the horizon radius at the infall time. At this timescale, the infalling bit also gets \textit{scrambled} with the 
black hole's interior. The act of an old black hole revealing what falls in later first 
is called \textit{information mirroring.} As far as we are aware, quantum information theory arguments alone cannot say much about how easily the Hayden-Preskill protocol of decoding the infalling information from the radiation can be done. An outstanding challenge is to come up with a controlled calculation in quantum gravity which captures these features of black hole dynamics. An elegant computational model of the Hayden-Preskill protocol has been discussed in \cite{Yoshida:2017non}.

Remarkable progress has been achieved recently via the use of the AdS/CFT dictionary in two-dimensional Jackiw-Teitelboim gravity \cite{Jackiw:1984je,Almheiri:2014cka} coupled to a conformal field theory. Employing the quantum extremal surface \cite{Engelhardt:2014gca} in the semiclassical evaporating black hole spacetime, the entanglement entropy of the Hawking radiation has been computed explicitly reproducing the Page curve in these simple quantum gravity setups \cite{Penington:2019npb,Almheiri:2019psf,Almheiri:2019hni}.  The geometric location of the quantum extremal surface post the Page time can be behind the black hole horizon. 
The Hayden-Preskill time for the encoding of the infalling information into the outgoing radiation is then captured precisely in the location of the quantum extremal surface. However, the mechanism of this encoding itself has not yet been addressed in this context.

\begin{figure}
\includegraphics[width = \linewidth]{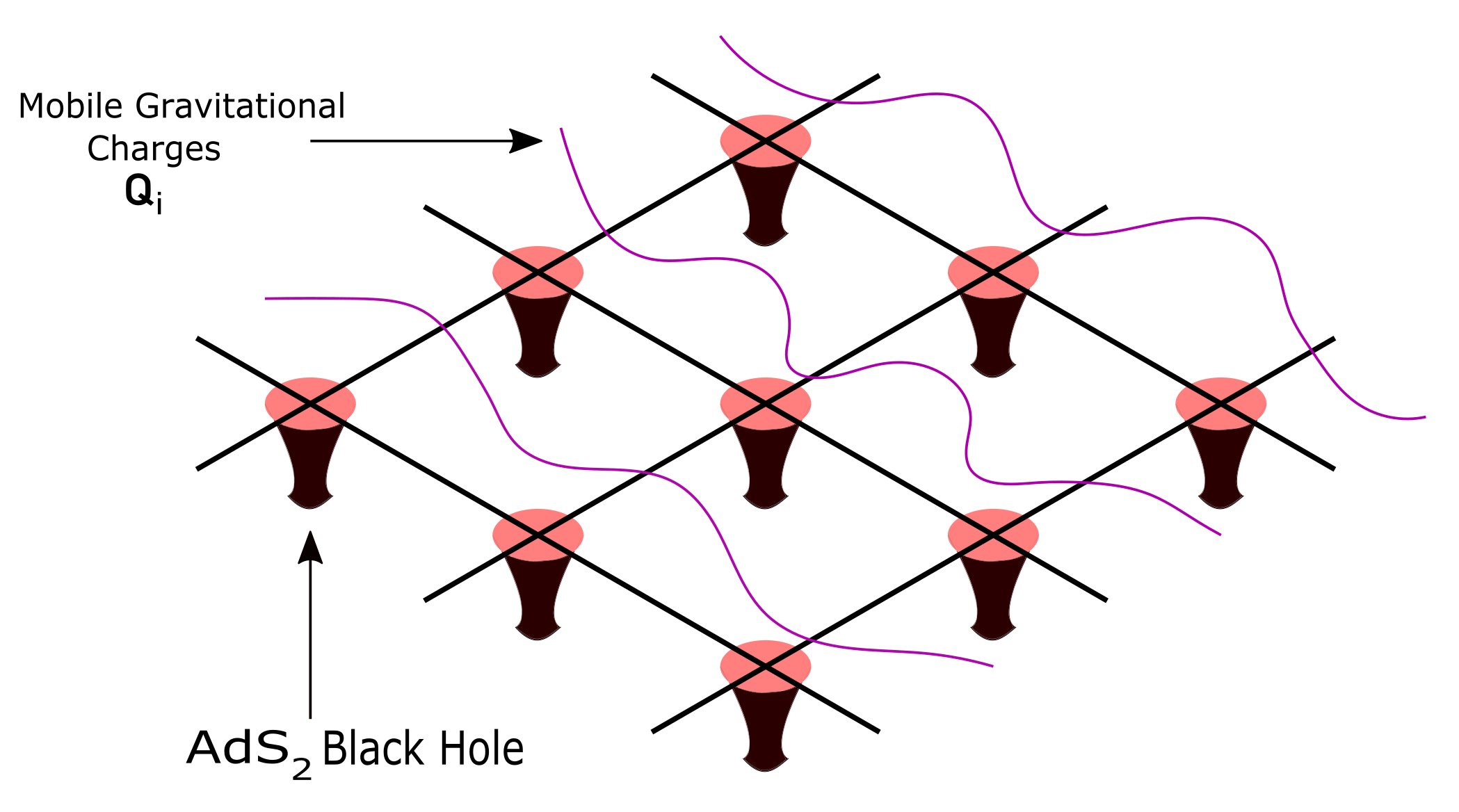}\caption{Schematic representation of our model, comprised of a lattice of $AdS_2$ black holes carrying $SL(2,R)$ charges $\bQ_i$ and interacting with mobile $SL(2,R)$ charges $\bq_i$ representing gravitational hair. Although we have shown a two-dimensional lattice in this figure, the lattice should actually have the dimensionality of the black hole horizon (in this case it represents the horizon of a four dimensional black hole). In this paper, we will study a chain with periodic boundary condition explicitly.}\label{Fig:OurModel}
\end{figure}

In this paper, with the motivation to understand the underlying mechanisms behind the information processing features of a higher dimensional black hole, particularly with respect to its scrambling and information mirroring properties, we construct a simple phenomenological model and study its dynamics. Our model is inspired by the instability of near-extremal black hole horizons to fragment into $AdS_2$ throats \cite{Brill:1991rw,Maldacena:1998uz}. This fragmentation instability is mediated by Brill instantons. 
Crucially a large number of low energy states appear at the boundaries of the instanton moduli space where two or more centers of the $AdS_2$ throats are mutually separated by sub-Planckian distances. Without attempting a first-principle derivation, we take a simple phenomenological approach to construct our model based on two ingredients: (i) a lattice of nearly $AdS_2$ throat geometries each described by Jackiw-Teitelboim (JT) gravity, (ii) quantum hair carrying $SL(2,R)$ charges that propagate over the lattice. A schematic representation can be seen in Fig. \ref{Fig:OurModel}. We find that our model can indeed be phenomenologically viable with a specific type of coupling between the $AdS_2$ throats and the gravitational hair. It captures the semi-classical features of a black hole, while also providing an explicit realization of  scrambling and information mirroring. The Hayden-Preskill protocol is  realized explicitly.

We study our model by taking the large $N$ limit simultaneously in all the $AdS_2$ throats in which the Hawking radiation becomes a semi-classical effect. We call this the semi-classical limit although the gravitational hair charges are described quantum-mechanically in this limit. Furthermore, if we consider coherent states of the hair quanta, we can also describe them via classical dynamics. In what follows, we show that a lot of the processing of the infalling information by the black hole can be understood from the interactions between the $AdS_2$ throats on the lattice and the quanta of the hair charges. The interaction of the Hawking quanta and the gravitational hair can be explicitly studied in our model, however we defer this to a future work. Although such interactions must be studied and are bound to provide further insights into quantum black hole dynamics via our model, they do not directly bear on the realization of the Hayden-Preskill protocol in the semi-classical limit. Since we focus primarily on the information processing mechanisms, we postpone the explicit computation of the Page curve in our model to the future. 

Our model captures only the near horizon dynamics. As such, it does not include the coupling of the hair to the asymptotic region of the full black hole geometry. In fact, we can easily argue that except for a monopole charge, the charges carried by coherent hair oscillations should not be conserved once we couple them to the asymptotic region. Such couplings can be included in our model without affecting any of the underlying mechanisms which stem from the interactions of the hair charges with the lattice of $AdS_2$ throats.

Briefly, our key results are as follows.
\begin{enumerate}
\item Our model provides an explicit realization of black hole microstates. These microstates can additionally support hair charge oscillations which can freely propagate without affecting them. 
\item The total energy in our model (conserved in the absence of perturbations) is simply the sum of the ADM masses of the $AdS_2$ throats giving the total black hole mass and the energy in the hair charges.
\item When any microstate with or without hair oscillations is perturbed by infalling matter in the form of shocks propagating through the $AdS_2$ throats, the full system quickly relaxes to another microstate with hair oscillations. 
\item Remarkably, the energy in the infalling matter is almost completely absorbed by the total black hole mass with the energy of the hair quanta remaining almost unaffected. Except for the moments of shock injection, the total black hole mass and the energy in the hair quanta are separately conserved to a very good approximation. Of course, the total energy is conserved exactly.
\item Finally, the hair charge oscillations that decouple from the final microstate carry the information about the locations and time-ordering of the infalling shocks. This leads to an explicit Hayden-Preskill protocol whose complexity scales only with that of the infalling bits and not of the underlying black hole dynamics. The information of the initial black hole state microstate is not retrievable easily although the dynamics is deterministic. A part of the hair charges gets locked in with the final black hole microstate.
\end{enumerate}

The plan of the paper is as follows. In Section \ref{Sec:Model}, we introduce our model and discuss analogies with the color glass condensate effective theory of saturated gluons in perturbative QCD. In Section \ref{Sec:Microstates}, we discuss the black hole microstate solutions and how they can support hair. We also show that the arrow of time emerges spontaneously in our model. In Section \ref{Sec:Shock}, we study the response of our model to perturbations in the form of infalling shocks and show that our model is phenomonologically viable. In Section \ref{Sec:HP}, we discuss the Hayden-Preskill protocol of retrieving the information of the shocks from the hair charge oscillations that decouple from the final black hole microstate solution. 
In Section \ref{Sec:Network}, we discuss some generalizations of our model in which we can replace the disconnected $AdS_2$ throats in the lattice by a network of $AdS_2$ throats connected via wormholes. We also discuss analogous models in condensed matter systems which can be built out of tensor network of a lattice of  SYK spin states (the latter states were first constructed in \cite{Kourkoulou:2017zaj}). Finally, in Section \ref{Sec:Conc}, we discuss some broader implications of our work.

\section{$SL(2,R)$ lattice model of the fragmented black hole horizon}\label{Sec:Model}

The fragmentation instability of a near-extremal compact $AdS_2 \times X$ black hole horizon refers to the instanton-mediated decay amplitude for the horizon to split into multiple $AdS_2 \times X$ throats \cite{Brill:1991rw,Maldacena:1998uz} (for a more recent discussion see \cite{Ooguri:2016pdq}). The instanton moduli space diverges when some of the throats produced by fragmentation become coincident. Presently we do not know how one can define a meaningful regularised amplitude for the fragmentation process. It is quite likely that there will be abundant new zero-energy states at the boundary of the instanton moduli space where these coincident limits are realized. These states can play a key role in understanding the quantum black hole.  We will refer to these gapless states as \textit{hair}\footnote{The role of soft hair in the resolution of the information loss paradox has been discussed in \cite{Hawking:2016msc} and elsewhere. Here we are dealing with a fragmented horizon unlike the discussion in these works. Although the mechanism is quite different in our model, it does seem that the hair can play an important role in retrieving infalling information as argued by these authors.}.

Motivated by this, we 
construct a simple model for the quantum black hole to study it as a quantum information processor. Here our goal is not to have a precise quantitative understanding of the black hole. Therefore, we will not worry whether our model reproduces the entropy of the quantum black hole except that in the semi-classical limit it should be proportional to the area in an obvious way. In fact, an exact model of a quantum black hole is stupendously ambitious with our present understanding of quantum gravity. As such, we will restrict ourselves to a simple construction that realizes the essence of the fragmented horizon picture and then investigate whether the model can be phenomenologically viable. 

The fragmented near-extremal horizon picture suggests a model in the form of a lattice of nearly-$AdS_2$ throats representing a fragmented (higher dimensional) space coupled to delocalized  gapless degrees of freedom representing the \textit{quantum hair}. The lattice has the same dimensionality as the black hole horizon. The key variables are
\begin{enumerate}
\item the immobile $SL(2,R)$ charges $\bQ_i$ of the nearly-$AdS_2$ throats described by JT gravity\footnote{{We expect that the  \textit{hard} degrees of freedom, namely the fragmented throats that accounts for almost all the energy (mass) of the black hole, should form a lattice just like the ions in a crystal. We do not account for disorder and other complications here. The fragmentation process implies that mass and other gravitational and non-gravitational charges of the near-horizon geometry will be split between the fragmented throats. We can also add non-gravitational charges such as electric and magnetic charges in our model. However, here we refrain from doing so for the sake of simplicity.}} at the corresponding lattice points $i$ and
\item the \textit{propagating} gapless $SL(2,R)$ charges $\bq_i$ that satisfy a source-free Klein-Gordon equation when they are decoupled from the immobile charges  $\bQ_i$.\footnote{One can think of these delocalized hair degrees of freedom as belonging to the yet-to-be fragmented near-extremal throat in the instanton-mediated fragmentation process. These hair should be naturally in the form of $SL(2,R)$ charges corresponding to the isometries of the unfragmented throat.}
\end{enumerate}
Let us first describe each of variables succinctly and then present our model which invokes a specific type of coupling between these variables.

\paragraph{The $SL(2,R)$ charges $\bQ_i$ and the corresponding $AdS_2$ throats:}The lattice $SL(2,R)$ charges $\bQ_i(u)$ can also be thought of as maps of the physical observer's time $u$ to a site-dependent time variable $t_i(u)$ corresponding to the ${\rm Diff}/SL(2,R)$ coset space of localised states. Thus each $SL(2,R)$ charge $\bQ_i(u)$ represent the time-dependent states of a SYK-type quantum dot system \cite{Sachdev:1992fk,Kitaev:2017awl} localized at the corresponding lattice site. 

We briefly review the unique map from $\bQ_i$ to $t_i(u)$. We also review how the localized state at the $i^{th}$ site is holographically dual to an $AdS_2$ black hole with time-dependent mass $M_i(u)$, which is proportional to the Schwarzian derivative of $t_i(u)$. 

It is useful to first define a \textit{standard local thermal time} $\tau_i(u)$ at each site using the relation
\be\label{Eq:t-tau-relation}
t_i(u) = \tanh\(\frac{\pi\tau_i(u)}{\beta}\).
\ee
If we perform the Euclidean rotation $t_i(u) \rightarrow i t_i(u)$, then $\tau_i(u) \rightarrow i \tau_i(u)$ and we readily see that the Euclidean $\tau_i(u)$ has period $\beta$. 
Concretely, the $SL(2,R)$ charges are related to (the Lorentzian) $t_i(u)$ via $\tau_i(u)$ as follows:
\ba\label{Eq:Q-T}
\Q^0_i &=& \frac{\beta}{2\pi}\(\frac{\tau_i'''}{{\tau_i'}^2}-\frac{{\tau_i''}^2}{{\tau_i'}^3}\)-\frac{2\pi}{\beta}\tau_i', \\
\Q^+_i &=& e^{\frac{2\pi\tau_i}{\beta}}\(\frac{\beta}{2\pi}\(\frac{\tau_i'''}{{\tau_i'}^2}-\frac{{\tau_i''}^2}{{\tau_i'}^3}\)-\frac{{\tau_i''}}{{\tau_i'}} \),  \\
\Q^-_i &=& e^{-\frac{2\pi\tau_i}{\beta}}\(\frac{\beta}{2\pi}\(\frac{\tau_i'''}{{\tau_i'}^2}-\frac{{\tau_i''}^2}{{\tau_i'}^3}\)+\frac{{\tau_i''}}{{\tau_i'}} \).
\ea
Above $'$ denotes the derivative w.r.t. the physical time $u$. To make the map $\bQ_i(u) \rightarrow \tau_i(u)$ 
unique, 
we can set the initial conditions at $u = 0$\footnote{We will consider the most general possibility in the next section. It will not be necessary that all these clocks are synchronized.}:
\ba\label{Eq:init}
\tau_i(u =0) = 0. 
\ea
To see this note that 
\ba\label{Eq:tau-p}
\tau_i' &=& \frac{\beta}{4\pi}\(\Q_i^-e^{\frac{2\pi\tau_i}{\beta}}+ \Q^+_i e^{-\frac{2\pi\tau_i}{\beta}}- 2 \Q_i^0\).
\ea
This differential equation has a unique solution for $\tau_i(u)$ with the given initial condition once we specify the charges $\bQ_i(u)$. In this discussion, $\beta$ is simply an arbitrarily chosen (site-independent) variable which will be useful for us later to simulate the system. Its physical significance, as we will see below, constitutes in simply choosing a reference state with temperature $\beta^{-1}$ with which we can define the time-reparametrizations that define the (localized correlation functions) of the lattice states. In fact, if we express the SL(2,R) charges in \eqref{Eq:Q-T} in terms of $t_i(u)$ instead of $\tau_i(u)$ by using \eqref{Eq:t-tau-relation}, we can explicitly see that $\beta$ disappears. 

For future reference, we also note that \eqref{Eq:Q-T} yields
\begin{align}\label{Eq:tau-pp-ppp}
\tau_i''&= \frac{\beta}{8\pi}\(\Q_i^-e^{\frac{2\pi\tau_i}{\beta}}+ \Q^+_i e^{-\frac{2\pi\tau_i}{\beta}}- 2 \Q_i^0\)\nonumber\\
&\(\Q_i^-e^{\frac{2\pi\tau_i}{\beta}}- \Q^+_i e^{-\frac{2\pi\tau_i}{\beta}}\), 
\nonumber\\
\tau_i'''&= \frac{\beta}{8\pi}\(\Q_i^-e^{\frac{2\pi\tau_i}{\beta}}+ \Q^+_i e^{-\frac{2\pi\tau_i}{\beta}}- 2 \Q_i^0\)
\nonumber \\ &\times\Big{(}({\Q_i^-}^2e^{\frac{4\pi\tau_i}{\beta}}+ {\Q^+_i}^2 e^{-\frac{4\pi\tau_i}{\beta}}\nonumber\\
&- \Q_i^0\(\Q_i^-e^{\frac{2\pi\tau_i}{\beta}}+ \Q^+_i e^{-\frac{2\pi\tau_i}{\beta}}\) \Big{)}.
\end{align}

Let us denote the $SL(2,R)$ invariant dot product of two charges $\boldsymbol{\mathcal{A}}$ and $\boldsymbol{\mathcal{B}}$ in the adjoint representation of $SL(2,R)$ as $\dt$, so that
\be\label{Eq:dot-product}
\boldsymbol{\mathcal{A}}\dt \boldsymbol{\mathcal{B}} := \mathcal{A}^0 \mathcal{B}^0 - \frac{1}{2}\( \mathcal{A}^+ \mathcal{B}^- +\mathcal{A}^- \mathcal{B}^+\).
\ee
We can readily see that the Casimir of the $SL(2,R)$ charge vector $\bQ_i$ is simply proportional to the Schwarzian derivative of $t_i(u)$ w.r.t. the physical time $u$. Explicitly,
\ba\label{Eq:Casimir}
\bQ_i\dt\bQ_i = - 2\, {\rm Sch}(t_i(u), u) = -2\,{\rm Sch}(\tau_i(u), u) + \frac{4\pi^2}{\beta^2} \tau_i'^2(u),
\ea
where we denote the Schwarzian as
\be
{\rm Sch}(f(u), u) := \frac{f'''(u)}{f'(u)} -\frac{3}{2}\(\frac{f''(u)}{f'(u)}\)^2.
\ee
For later convenience, we define
\be
{\rm Sch}_i(u) :={\rm Sch}(t_i(u), u).
\ee
One can also verify these useful identities, which show that the $SL(2,R)$ charges $\bQ_i$ must be constant if the Schwarzian derivatives of $t_i(u)$ w.r.t. $u$ are constant:
\ba\label{Eq:Q-ids}
{\Q_i^0}' = \frac{\beta}{2\pi}\frac{{\rm Sch}_i'}{\tau_i'}, \quad {\Q_i^\pm}' = \frac{\beta}{2\pi}e^{\pm\frac{2\pi}{\beta}\tau_i}\frac{{\rm Sch}_i'}{\tau_i'}.
\ea

If JT-gravity provides the dual gravitational description \cite{Almheiri:2014cka,Jensen:2016pah,Engelsoy:2016xyb,Maldacena:2016upp}, then the holographic dual of the localized state at the $i^{th}$ site can be represented as an $AdS_2$ black hole with a time-dependent mass $M_i(u)$ and thus also a time-dependent horizon $r_s(u)$ without loss of generality. In the ingoing Eddington-Finkelstein gauge, the corresponding metric takes the form
\be\label{Eq:EFMu}
{\rm d}s^2 = - \frac{2}{r^2} {\rm d}r{\rm d}u -\(\frac{1}{r^2}-\frac{1}{r_s^2(u)}\){\rm d}u^2.
\ee
 Above, we have dropped the site index $i$. The $ADM$ mass of this $AdS_2$ black hole is
 \begin{equation}\label{Eq:M-rs}
 M(u) = \kappa \frac{1}{r_s^2(u)},
 \end{equation}
 where $\kappa$ is a site-independent parameter with mass dimension minus one. We will treat $\kappa$ simply as a parameter of our phenomenological model. 
 We will reserve a description of the full JT gravity solution for Section \ref{Sec:Shock}.
 
 It remains to identify $M_i(u)$ with the corresponding $t_i(u)$ which identifies the dual localized state in the coset space ${\rm Diff}/SL(2,R)$. To see this explicitly, we need to perform the diffeomorphism
\ba\label{Eq:Diffeo}
t = t(u), \quad \rho = \frac{t'(u) r}{1- \frac{t''(u)}{t'(u)}r},
\ea
 which preserves the Eddington-Finkelstein gauge, but maps the metric \eqref{Eq:EFMu} to the pure $AdS_2$ vacuum
\be\label{Eq:EFM0}
{\rm d}s^2 = - \frac{2}{\rho^2} {\rm d}\rho{\rm d}t -\frac{1}{\rho^2}{\rm d}t^2,
\ee
provided
\be\label{Eq:M-Sch}
\frac{1}{r_s^2(u)} = - 2\, {\rm Sch}(t(u),u).
\ee
Similarly, replacing $t(u)$ by $\tau(u)$ in \eqref{Eq:Diffeo} we find that the metric \eqref{Eq:EFMu} dual to the localized state maps to the metric of an $AdS_2$ black hole with $1/r_s^2 = 4\pi^2/\beta^2$ and the corresponding Euclidean black hole has time period $\beta$. Also note that if $\tau(u) = u$, then \eqref{Eq:Q-T} implies that $\Q^0 = - 2\pi/\beta$ and $\Q^\pm = 0$ so that the Casimir is $4\pi^2/\beta^2$. Therefore, $\tau(u)$ indeed is the map to the black hole with fixed mass $4\kappa\pi^2/\beta^2$.

To summarize, the $SL(2,R)$ lattice charges $\bQ_i(u)$  represent  $AdS_2$ black holes at the corresponding sites with masses
\be
M_i(u) =  - 2 \kappa\, {\rm Sch}(t_i(u),u) = \kappa\bQ_i(u) \dt \bQ_i(u), 
\ee 
as seen in \eqref{Eq:M-Sch} and \eqref{Eq:M-rs}. Also $t_i(u)$ represents the dual localized states at corresponding sites belonging to the coset space ${\rm Diff}/SL(2,R)$. The JT gravity descriptions elegantly portray $t_i(u)$ as the map of the physical time $u$ shared by all the localized states to the time $t_i$ of the vacuum state of the theory at $i^{th}$ site. The  intra-site correlation functions of the localized states are simply given by the conformal transformations $u \rightarrow t_i(u)$ of the corresponding vacuum correlation functions.

\paragraph{The quantum hair:} At each site we additionally have the quantized $SL(2,R)$ hair charges $\bq_i$ which are propagating gapless excitations on the lattice. When we switch off the coupling to the localized lattice charges $\bQ_i$, we simply assume that the quantum hair charges follow the discrete version of the Klein-Gordon equation
\ba
\bq_i'' - \frac{1}{\si^2}\( \bq_{i-1} + \bq_{i+1} - 2 \bq_i\) = 0.
\ea
Above $\si$ has mass dimension one.

Although the $\bq_i$ are quantum variables, we can consider coherent states of these hair quanta so that they can be thought of as classical variables for practical purposes.

\paragraph{Coupling lattice charges to the quantum hair:} Our model must involve a coupling between the $SL(2,R)$ lattice charges and the quantum hair. This coupling should be such that at least these three basic requirements are satisfied:
\begin{enumerate}
\item The full dynamics has only one overall $SL(2,R)$ symmetry and not $SL(2,R) \otimes SL(2,R) \otimes \cdots \otimes SL(2,R)$ symmetry with $n$ factors where $n$ corresponds to  the number of lattice sites. This global $SL(2,R)$ symmetry should correspond to that of the isometry of the classical (unfragmented) near-extremal horizon.
\item Causality must hold. In particular, if a specific site $i$ is shocked by an infalling bit then $t_j'''(u)$ cannot change before the time $u = L\vert j -i\vert$, where $L$ is the lattice spacing. (Note the equation of motion of $t_i(u)$ should be fourth order for it to be global $SL(2,R)$ invariant.)
\item The full system should have a conserved energy.
\end{enumerate}

These criteria are insufficient to give us a unique model. Therefore we will be guided by phenomenology. It turns out that the simplest model with the desired phenomenological features can be described by these equations:
\ba\label{Eq:Model}
M_i' &= &- \la\(\bQ_{i-1}+ \bQ_{i+1}-2 \bQ_i\)\dt \bq_i',\nonumber\\
\bq_i'' &=&\frac{1}{\si^2}\( \bq_{i-1} + \bq_{i+1} - 2 \bq_i\) \nonumber\\
&+&  \frac{1}{ \la^2}\(\bQ_{i-1}+ \bQ_{i+1}-2 \bQ_i\).
\ea
Above $\lambda $ is the parameter for the coupling between the immobile lattice charges and the mobile gravitational charges. It has the same mass dimension as $\sigma$.  These equations should be viewed as the dynamical equations that determine the evolution of the lattice variables $t_i(u)$ and the $\bq_i(u)$. We will later 
show that $\lambda$ must be positive. In order to obtain the continuum limit of our model, we need to take the limit $n$, the number of lattice sites, to infinity with $n\bQ_i$, $n\bq_i$, $nM_i$ and $n\kappa$ kept finite.

{We can consider further generalizations in which we can add higher derivative terms to the equation of motion for $\bq_i$  (along with necessary higher order interactions with $\bQ_i$) and also introduce higher-spin fields in the JT gravity theories describing each throat. These can model stringy effects. Furthermore, such modifications can be made such that we generate black hole microstates (discussed in the following section) with the same features. We will however not concern ourselves with such stringy effects in the present work.} 

We readily see that the model has a conserved total energy $\me$ given by
\begin{align}\label{Eq:E}
\me &=  \sum_i M_i + \frac{\lambda^3}{2}\sum_i \bq'_i \dt \bq'_i \nonumber\\
&+ \frac{\lambda^3}{2\sigma^2}\sum_i \(\bq_{i+1} -\bq_{i}\) \dt\(\bq_{i+1} -\bq_{i}\)\\\nonumber
&=  \kappa \sum_i \bQ_i\cdot \bQ_i + \frac{\lambda^3}{2}\sum_i \bq'_i \dt \bq'_i \nonumber\\
&+ \frac{\lambda^3}{2\sigma^2}\sum_i \(\bq_{i+1} -\bq_{i}\) \dt\(\bq_{i+1} -\bq_{i}\).
\end{align}
It is easy to see that our model retains a global $SL(2,R)$ symmetry which is also the isometry of the original non-fragmented near-extremal black hole horizon.

Our model can be compared with color glass condensate effective theory of saturation physics in QCD \cite{Gelis:2010nm}. This theory is an effective description for low $x-$gluons in a hadron where $x$ denotes the fraction of longitudinal momentum carried by the gluon. Such gluons can be described by classical chromo-electromagnetic flux tubes with transverse widths of the order of $Q_s^{-1}(x)$. The saturation scale $Q_s(x)$ is the transverse virtual momentum energy scale for a given $x$ below which the gluons have occupation numbers of the order of $1/\alpha_s(Q)$, the inverse of the strong-coupling constant. Thus for $x < x_c$, where $x_c$ is a suitably chosen cut-off, one may use classical Yang-Mills equations to describe these \textit{saturated} gluons.  These field equations however must have color sources generated by the $x > x_c$ gluons which are frozen (static) at the time-scales of the effective theory and have a scale-dependent stochastic distribution. This effective description of low $x-$gluons is valid only at weak coupling and is applicable to the initial stages of heavy-ion collisions.

In our model, we can think of the $AdS_2$ throats as the analogues of the chromo-electromagnetic flux tubes which admit classical description. Here we are assuming a large $N$ limit in the dual SYK-type quantum dots which allows us to neglect quantum corrections in each of the holographic JT gravity descriptions. The analogues of the high energy gluons are the mobile gravitational $SL(2,R)$ hair charges. The crucial difference is that the hair charges are time-dependent while the color charges of the high $x-$gluons  are frozen in the effective description. In fact accounting for red-shift immediately tells us that the dynamics of the interior of the $AdS_2$ throats will appear to be frozen to the physical observer. This naturally fits into the UV/IR type duality in the traditional AdS/CFT correspondence.\footnote{In AdS/CFT, it is natural that red-shift freezes the near-horizon degrees of freedom from the viewpoint of the observer at the boundary. These describe the IR of the dual theory. However, in the dual theory the high momentum gluons will be naturally frozen due to time-dilation from the point of view of the low energy theory.} 

 One may 
 note a further analogy here. The color glass condensate effective theory has a natural stochastic element, namely the color source distribution of the high $x-$gluons.  Recently it has been demonstrated that JT gravity is dual to an ensemble of quantum-mechanical Hamiltonians \cite{Saad:2019lba,Stanford:2019vob,Witten:2020bvl}. In fact the SYK model \cite{Sachdev:1992fk,Kitaev:2017awl} itself involves stochastic averaging over four-Majorana fermion couplings. Our model is 
essentially 
the mobile gravitational hair $\bq_i$ interacting with an ensemble of Hamiltonians dual to the JT gravity at each lattice site. In a way our model of a near-extremal quantum black hole is the strong coupling holographic analogue of saturation phenomena in perturbative QCD except that the roles of the UV and IR degrees of freedom are inverted in terms of which provide stochastic sources for the evolution of the other.\footnote{At strong coupling and large $N$, it is natural that the saturation scale will not be very separated from an effective thermal scale. Large density accumulation in gravity is expected to trigger formation of black hole microstates.} The IR in the form of JT gravities provide stochastic sources (via their holographic interpretation) for the field equations of the UV degrees of freedom described by the hair $\bq_i$ rather than the other way round.  Unlike a traditional holographic description, our effective description involves a fragmented spacetime and can be naturally a part of non-supersymmetric version of AdS/CFT with fragmentation instabilities. Our discussion here is admittedly heuristic. The actual physical description of a near-extremal quantum black hole is likely to involve a more complex generalization of our simple phenomenological model.

 \section{The quantum black hole microstates: hairy and bald}\label{Sec:Microstates}
 We need to first identify the solutions of our model which can be naturally interpreted as the microscopic states of the quantum black hole. We should rather call these black hole mesoscopic states because we have only a coarse-grained description in terms of the lattice charges $\bQ_i$, or equivalently only the time-reparametrization modes $t_i(u)$ for the SYK-type quantum dots on the lattice sites. Nevertheless, we will still use the word microstates to denote these solutions. Before presenting the phenomenological features which justify the model, it will be useful for us to understand these general solutions representing the microscopic states. We will show that each microstate can support hair in the form of decoupled quanta of the $\bq_i$ charges manifesting as coherent oscillations which freely propagate over the lattice without affecting it.
 
 For the purpose of this discussion, it is useful to first decompose the full conserved energy $\me$ in \eqref{Eq:E}  into two parts as below
 \ba
& \me = \me_{\Q}  + \me_{\q}, \quad\me_{\Q} = \sum_i M_i, \\\nonumber &\me_{\q} =  \frac{\lambda^3}{2}\sum_i \bq'_i \dt \bq'_i + \frac{\lambda^3}{2\sigma^2}\sum_i \(\bq_{i+1} -\bq_{i}\) \dt\(\bq_{i+1} -\bq_{i}\).
 \ea
 
 The solutions of \eqref{Eq:Model} 
 have the following properties:
 \begin{enumerate}
 \item The  masses $M_i$ and the lattice charges $\bQ_i$ are time-independent. 
 \item The hair charges $\q_i$ are a linear superposition of two parts: (i) a static part which is \textit{locked in} with the $\bQ_i$ lattice charges, and (ii) a monopole (homogeneous) part which is decoupled from $\bQ_i$. Crucially, both $\me_{\Q}$ and $\me_{\q}$ should be separately time-independent.
 \end{enumerate}
 Hair charge modes which are not monopoles (i.e. homogeneous) can be expected to be \textit{carried away to asymptotic infinity} via gravitational radiation and Hawking quanta radiated from the $AdS_2$ throats which also interact with the hair charges. Although we have not included any type of coupling of our model of the quantum black hole to the asymptotic region of spacetime or considered quantum gravity effects in the $AdS_2$ throats (particularly the Hawking radiation), we should expect that only the monopole term can \textit{remain with the black hole.} A microstate solution with hair will refer to a solution with additional coherent states of $\q_i$ that are propagating freely in the background of a particular microstate  solution without affecting it.

To find microstate solutions we recall the identites \eqref{Eq:Q-ids} which imply that for the $\bQ_i'$ to all vanish, it is sufficient that  $M_i'$ all vanish. Then we see from \eqref{Eq:Model} that for $M_i'$ to vanish we need  $\bq_i'$ to be parallel to each other, i.e.  $$ \bq_i' = q_i' \boldsymbol{\xi}$$ where $\boldsymbol\xi$ is a constant site-independent $SL(2,R)$ charge vector of unit norm and also 
$$ \boldsymbol{\xi} \cdot \bQ_i = Q$$ 
 with $Q$ a constant that does not depend on the lattice site. Since our model has a global $SL(2,R)$ symmetry, we can always choose $\xi^\pm = 0$ and $\xi^0 = 1$ without loss of generality by a global $SL(2,R)$ rotation.\footnote{Note this also exhausts our utilization of the global $SL(2,R)$ symmetry. If the throats were connected by an unfragmented spacetime such as in the case of the eternal black hole with two asymptotic boundaries, then the choice of a global (Kruskal) time coordinate would have achieved the same effect of exhausting the overall $SL(2,R)$ freedom. In our case, the choice of the direction of the homogeneous component of the lattice $SL(2,R)$ charges is actually related to how we glue our model to the asymptotic unfragmented geometry. Of course, this has no bearing on the dynamics of our model. We thank Gautam Mandal for a discussion on this issue. } Therefore, this implies that 
 \be\label{Eq:q0peq}
 \q^{0\prime}_i  = q_i', \quad  \q^{\pm\prime}_i  = 0
 \ee  and 
 \be\label{Eq:Q0eq}
 \Q^0_i  = Q.
 \ee
 Note that although we need $\Q^0_i$ to take the same values at all lattice points for all $\bQ_i'$  to  vanish, $\Q^\pm_i$ can take arbitrary constant values. This implies that the $AdS_2$ black holes in the lattice can have different (constant) masses $M_i$. The $SL(2,R)$ charge vectors describing the black hole interior are thus similar to disordered ferromagnets. While one of the components of these charge vectors are homogeneous representing order, the other components are stochastic and uncorrelated. We will soon see that the consequence of having $\Q^0_i = Q$ will be that the arrows of time of the lattice sites given by ${\rm sgn}(t'_i(u))$ will be the same for all lattice sites.
 
 We also see from the second equation of  \eqref{Eq:Model} that all $\bq''_i$ vanish if
 \be\label{Eq:qloc}
 \bq_i  = \bq_i^{loc} := -\frac{\si^2}{\la^2} \bQ_i+ \boldsymbol{\mathcal{K}}
 \ee
The hair charges $ \bq_i^{loc}$ are thus locked in with the $\bQ_i$. Note $\boldsymbol{\mathcal{K}}$ has no lattice site index. Since $\Q^0_i = Q$,  it follows that $\q^0_i = -(\si^2/\la^2) Q + \mathcal{K}^0$ and are thus homogeneous too. However, we note that \eqref{Eq:q0peq} and \eqref{Eq:Q0eq} also allow a monopole term such that 
\be
q'_i= \alpha,
\ee
with $\alpha$ being a site-independent constant implying
\be\label{Eq:qmon}
\bq_i = \bq_i^{mon}, \quad (\q_i^{mon})^0 = \alpha\,u, \quad (\q_i^{mon})^\pm = 0.
\ee
The masses $M_i$ and the $\bQ_i$  are then constant with \eqref{Eq:Q0eq} satisfied while the hair charges are
\be\label{Eq:qmicro}
\bq_i = \bq_i^{loc} + \bq_i^{mon}.
\ee
 The energy in the hair charges $\me_{\q}$ is then given by the sum of two parts
\ba\label{Eq:Eq-split-1}
&\me_{\q} = \me_{\q}^{pot} + \me_{\q}^{mon}, \nonumber\\
&  \me_{\q}^{pot} =-\frac{\si^2}{2\la}\sum_i \(\Q^{+}_{i+1} -\Q^{+}_{i}\) \(\Q^-_{i+1} -\Q^-_{i}\),
\nonumber\\ &
\me_{\q}^{mon} = \frac{1}{2}\la^3 \alpha^2.
\ea 
Note $\me_{\q}^{pot}$ arises from $\bq_i^{loc}$ and $\me_{\q}^{mon}$ is due to the monopole $\bq_i^{mon}$.

To summarize, the microstate solutions are those with the following configuration:
\begin{enumerate}
\item Constant and homogeneous $\Q^0_i = Q$,
\item Constant values of $\Q^\pm_i$, so that the masses of the $AdS_2$ black holes are
\be M_i = \kappa(Q^2 - \Q^+_i\Q^-_i)\ee
We will set $\kappa = 1$ from now on.
\item Hair charges which can be decomposed into two pieces $\bq_i = \bq_i^{loc} + \bq_i^{mon}$ where $\bq_i^{loc}$ is locked in with $\Q^0_i$ as in \eqref{Eq:qloc} and $ \bq_i^{mon}$ is the monopole term \eqref{Eq:qmon}.
\end{enumerate}

It is also easy to see that since  $\bQ_i'$ vanish and the equation for $\bq_i$ in \eqref{Eq:Model} becomes source-free by virtue of \eqref{Eq:q0peq} and \eqref{Eq:Q0eq}, each microstate can support hair oscillations of the form \be\label{Eq:qrad}\bq_i = \bq^{rad}_i, \quad (\bq^{rad}_i)^\pm = 0, \quad (\bq^{rad}_i)^0 = q_i\ee with $q_i$ satisfying the  free lattice Klein-Gordon equation:
 \be\label{Eq:KGj}
 q_i'' = \frac{1}{\sigma^2} \(q_{i+1} + q_{i-1}- 2 q_i \).
 \ee
 We will call this the radiation component of the hair. These coherent oscillations should exclude the monopole term to avoid double counting, so that
 \be\label{Eq:qrad2} \sum_i q_i' = \sum_i q_i = 0\ee for all time. (Note \eqref{Eq:KGj} implies $\sum_i q_i'$ is constant which can be set to zero.) These can freely propagate on any microstate solution background without affecting the latter. Adding such coherent oscillations, we get \be \bq_i = \bq_i^{loc} + \bq_i^{mon} + \bq^{rad}_i \ee and such solutions are then microstate solutions with hair.
 
 It is easy to see that for microstate solutions with hair
 \begin{align}\label{Eq:Eq-split-2}
\me_{\q} &= \me_{\q}^{pot} + \me_{\q}^{mon} + \me_{\q}^{rad}, \nonumber\\
\me_{\q}^{pot} &=-\frac{\sigma^2}{2\lambda}\sum_i \(\Q^{+}_{i+1} -\Q^{+}_{i}\) \(\Q^-_{i+1} -\Q^-_{i}\),
\nonumber\\ 
\me_{\q}^{mon} &= \frac{1}{2}\la^3 \alpha^2, \nonumber\\
 \me_{\q}^{rad}&=  \frac{\la^3}{2}\sum_i {q'_i}^2 + \frac{\la^3}{2\si^2}\sum_i \(q_{i+1} -q_{i}\)^2.
\end{align}
 Clearly, if $\lambda > 0$, then
 \be
 \me_{\q}^{rad} \geq 0, \quad \me_{\q}^{mon} \geq 0.
 \ee
 This leads us to set $\lambda > 0$. The restriction $M_i \geq 0$ should be be imposed so that $\me_{\Q} \geq 0$ also. Only the $\me_{\q}^{pot}$ term in the energy can be negative. However since $\Q^\pm_i$ are randomly chosen, averaging over microstates should give $\me_{\q}^{pot} = 0$. Therefore, the average total energy in the ensemble of microstate solutions is positive definite. It is to be noted that the split of $\me_{\q}$ into the three terms $\me_{\q}^{pot}$, $\me_{\q}^{mon}$ and $\me_{\q}^{rad}$ makes sense only for the microstate solution with(out) hair and not generally. 
 
{We can readily prove that the microstate solutions discussed here are unique. Let us discuss the hairless solutions first. In order to produce stationary solutions we need the right hand side of the second equation of our model \eqref{Eq:Model} to vanish. The latter are linear equations for $\bq_i$ and therefore \eqref{Eq:qloc} are the unique solutions.  In order to preserve the conditions \eqref{Eq:qloc} we need the components of $\bq_i'$ to be vanishing in any given (internal) direction, or to be homogeneous and constant.  Choosing an appropriate global $SL(2,R)$ frame, the non-vanishing homogeneous components should satisfy \eqref{Eq:q0peq} without loss of generality. The rest of the construction, including the addition of hair oscillations, then follows.}
 
We can readily note that for physical viability, the lattice charges $\bQ_i$ in the black hole microstate solutions need to satisfy further restrictions. We 
need $t_i(u)$ and therefore $\tau_i(u)$ to be real at each site in the entire range of the physical observer's time, $-\infty < u < \infty$. It is also necessary that $t_i(u)$ and therefore $\tau_i(u)$ are sufficiently smooth (these and their first and second derivatives must be continuous) as dictated by our equations \eqref{Eq:Model}. Furthermore we will also require that we can synchonize the clocks at the lattice sites at time $u=0$ via \eqref{Eq:init}. The justification is that there is a preferred global time-coordinate in which we can do this synchronization, which is the moment when the quantum black hole is \textit{formed}. However, this is not crucial for what follows and we can readily generalize our initial conditions \eqref{Eq:init} to 
\be\label{Eq:init-2}
\tau_i(u = 0) = \tau_{i0}.
\ee
Once we perturb an initial microstate satisfying the initial conditions \eqref{Eq:init} for $\tau_i(u)$, the final microstate to which the system will relax to will satisfy the above initial conditions for $\tau_i(u)$ actually.

Setting $\Q^0_i = Q$ and using \eqref{Eq:tau-p} and \eqref{Eq:init} we readily see that the solutions for $\tau_i(u)$ given the data for constant $Q$ and $\Q^\pm_i$ in these black hole microstate solutions  solutions are as follows.  If $\Q^-_i  \neq 0$ and $M_i > 0$, then
\begin{align}\label{Eq:tauiu}
\tau_i(u)  &= \frac{\beta}{2\pi}\log\Big[ \frac{Q}{\Q^-_i}\nonumber\\
&-\frac{\sqrt{M_i}}{\Q^-_i}
\tanh\(\frac{\sqrt{M_i}}{2}u + {\rm arctanh}\(\frac{Q- \Q^-_i}{\sqrt{M_i}}\)\)\Big]
\end{align}
and
\begin{align}\label{Eq:tiu}
t_i(u)  &= \tanh \Big{(}\frac{1}{2}\log\Big[ \frac{Q}{\Q^-_i}\nonumber\\
&-\frac{\sqrt{M_i}}{\Q^-_i}
\tanh\(\frac{\sqrt{M_i}}{2}u + {\rm arctanh}\(\frac{Q- \Q^-_i}{\sqrt{M_i}}\)\)\Big]\Big{)}.
\end{align}
If $\Q^-_i  = 0$ , then 
\ba\label{Eq:tauiu-2}
\tau_i(u)  = \frac{\beta}{2\pi}\log\left[ \(\frac{2 Q -\Q^+_i}{2 Q }\) e^{-Q u}+\frac{\Q^+_i}{2 Q}\right]
\ea
 and
 \ba\label{Eq:tiu-2}
t_i(u)  = \tanh\(\frac{1}{2}\log\left[ \(\frac{2 Q -\Q^+_i}{2 Q }\) e^{-Q u}+\frac{\Q^+_i}{2 Q}\right]\).
\ea
Finally, if $M_i  = 0$ and $\Q^0_i =  Q = \Q^-_i/\rho_i = \Q^+_i \rho_i $, then 
\ba
\tau_i(u) = \frac{\beta}{2\pi}\log\left[\frac{1}{\rho_i} +\frac{\frac{2}{\rho_i}- 2}{Qu(\rho_i -1) -2} \right], \nonumber\\ t_i(u) = \tanh\(\frac{1}{2}\log\left[\frac{1}{\rho_i} +\frac{\frac{2}{\rho_i}- 2}{Qu(\rho_i -1) -2} \right]\).
\ea
 Note for $\rho_i = 1$, the above reduces to $\tau_i(u)= t_i(u)  = 0$.

Requiring $\tau_i(u)$ and $t_i(u)$ to be real and have continuous first and second derivatives for $-\infty < u < \infty$ 
implies two possibilities for $M_i \geq 0$. The first is that
\be\label{Eq:1poss}
Q  \leq - M _i, \quad  \Q_i^\pm \leq 0, \quad \Q_i^+ + \Q_i^- \geq 2 Q.
\ee 
The second possibility is that
\be\label{Eq:2poss}
Q  \geq M _i, \quad  \Q_i^\pm \geq 0, \quad \Q_i^+ + \Q_i^- \leq 2 Q.
\ee 
Remarkably the first possibility implies that $\tau'_i \geq 0$, while the second implies that $\tau'_i \leq 0$ for all $i$ and for $-\infty < u < \infty$. Thus, there has to be a strict arrow of time at every site in our model  \eqref{Eq:Model} and all these arrows of time have to be aligned in the same direction (future/past) in the black hole microstate solutions with(out) hair.  \textit{A global (uniform) arrow of time is therefore emergent in our model}. Given that we have fractionalized space into several quantum dots spread over a lattice and that we have an independent time $t_i(u)$ for each lattice state, it is reassuring to see this feature. We note that it is important to invoke the positivity of energy that requires $M_i \geq 0$ for this global arrow of time to emerge.

Our conclusions do not change for more general initial conditions \eqref{Eq:init-2}. In \eqref{Eq:tauiu}, etc. we simply need to replace $u$ by $u - v_{i}$ with a suitably chosen constant $v_{i}$ for the corresponding site.

\begin{figure}
\includegraphics[width=\linewidth]{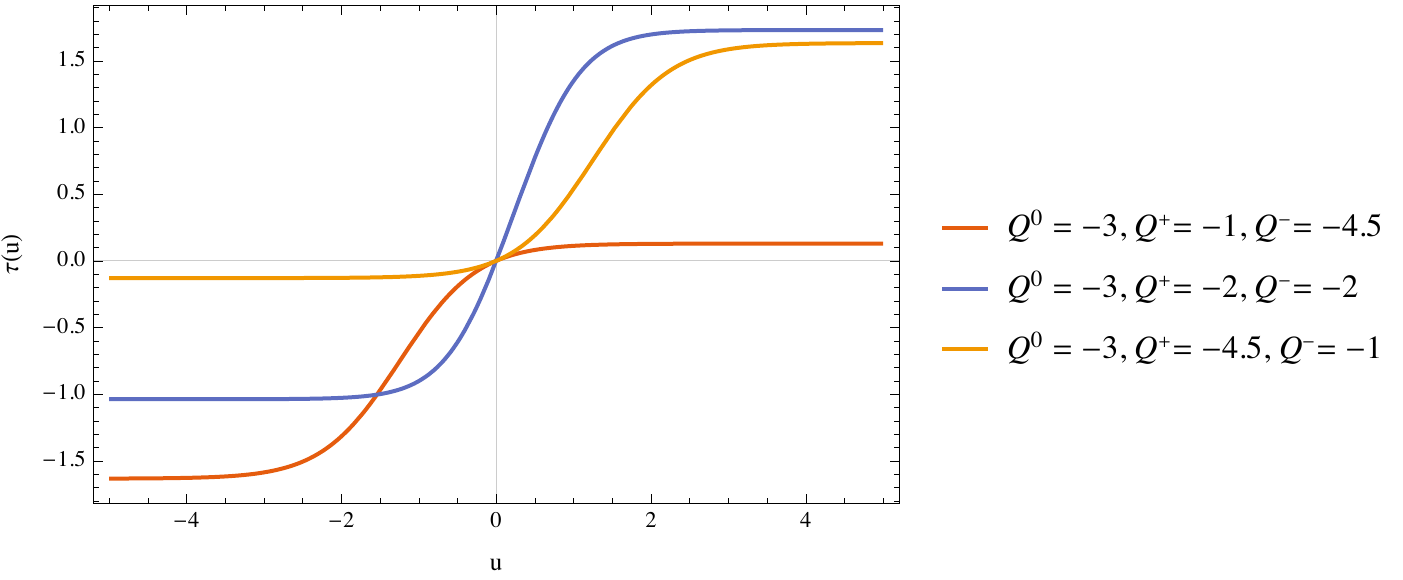}\caption{Plot of $\tau(u)$ vs $u$ given by \eqref{Eq:tauiu} for various choices of values of the $SL(2,R)$ charges allowed by \eqref{Eq:1poss} and $\beta = 2\pi$.}\label{Fig:tauplot}
\end{figure}

We will choose the first possibility \eqref{Eq:1poss} since we want our global arrow of time to point towards the future. This implies the following parametrization of the $SL(2,R)$ lattice charges with $Q > 0$ and $0 \leq M_i \leq Q$ (note we have redefined $Q$ below with a minus sign for future convenience):
\ba\label{Eq:Q-par}
&\Q^0_i = - Q, \quad \Q^+ = -\rho_i\sqrt{Q^2 - M_i}, \nonumber\\
&\Q^- = -\frac{\sqrt{Q^2 - M_i}}{\rho_i}
\ea
such that
\be\label{Eq:rho-range}
\sqrt{\frac{\sqrt{ Q } - M_i }{\sqrt{Q } + M_i}}\leq\rho_i \leq \sqrt{\frac{\sqrt{ Q } + M_i }{\sqrt{ Q } - M_i}}.
\ee
We note that when $Q$ approaches $M_i$, the range of possible values of $\rho_i$ extends over the entire positive real axis. On the other hand, when $Q$ is very large, the possible range of values of $\rho_i$ extends only over a tiny interval which collapses to $1$ eventually. For examples of $\tau(u)$, see Fig. \ref{Fig:tauplot}.
It is natural to associate a black hole macrostate with a given value of the total mass $\me_{\Q} := M$ and the order parameter $Q$. The counting of the microstates mainly involves taking into account the partitioning of $M$ into $M_i$ subject to the restriction $0 \leq M_i \leq Q$, as well as the possible values of $\bQ_i$ 
as in \eqref{Eq:Q-par} with each $\rho_i$ subject to the restriction given by \eqref{Eq:rho-range}.

A pertinent question is whether a hairy microstate with lattice charges parametrized via \eqref{Eq:Q-par} subject to the restrictions \eqref{Eq:rho-range} and $0 \leq M_i \leq Q$ will relax to another hairy microstate with lattice charges subject to these same restrictions after a perturbation. This is indeed borne out -- we will discuss this soon.

Finally, we note that $\tau_i(u)$ is physically the map of the time of the lattice state to that of an $AdS_2$ black hole with mass $M= 4\pi^2/\beta^2$. Since this is an arbitrarily chosen standard black hole clock we can set $\beta = 2\pi$ so that this standard black hole has unit mass. We will discuss the full JT-gravity solutions in the next section.

 \section{Shocks reveal phenomenological viability}\label{Sec:Shock}
 
 The obvious question is whether  our $SL(2,R)$ lattice model walks and talks like a black hole. We can specifically ask what happens if one or more compact objects fall(s) into one of our typical black hole microstate solutions. Since we have fragmented the horizon geometry into a lattice of $AdS_2$ throats, compact infalling objects are  simply infalling shocks in these throats.  The masses and the $SL(2,R)$ charges of the corresponding throats jump instantaneously at these moments of shock injections. For phenomenological viability, we will expect the following features:
 \begin{enumerate}
 \item After all the shocks fall into the initial black hole microstate solution with or without decoupled hair charge oscillations, the full system should relax to another such black hole microstate eventually.
 \item The energy injected via the shocks should almost go fully towards increasing the overall mass $M := \me_{\Q}$. 
 \item During the course of time-evolution, the sum of the black hole masses $\me_{\Q}$ and the energy in the gravitational hair charges $\me_{\q}$ should be conserved separately to a very good approximation.
 \end{enumerate}
 These features, if validated, will ensure that if we are looking at the system from \textit{afar}, it will seem to behave like an ordinary black hole particularly when we are taking the large $N$ limit at each lattice point simultaneously. Note none of these features are built into the construction of our model -- these have to emerge from the equations of motion. Nevertheless, both the initial and final microstate solutions will have inhomogeneities at sub-horizon scale.

 The equations governing the full system in the presence of shocks take the form:
 \ba\label{Eq:Model-Shocked}
M_i' &= &- \la\(\bQ_{i-1}+ \bQ_{i+1}-2 \bQ_i\)\dt \bq_i' + \sum_A e_{i,A} \delta(u - u_{i,A}),\nonumber\\
\bq_i'' &=&\frac{1}{\si^2}\( \bq_{i-1} + \bq_{i+1} - 2 \bq_i\) \nonumber\\
&+&  \frac{1}{ \la^2}\(\bQ_{i-1}+ \bQ_{i+1}-2 \bQ_i\),
\ea
 where $e_{i,A}$ is the amount of energy injected into the $i^{th}$ throat at $u = u_{i,A}$. We refer the reader to Appendix \ref{App:Dilaton} for the derivation of the solution of the JT-gravity in each $AdS_2$ throat. A brief summary is as follows. In the locally $AdS_2$ geometry \eqref{Eq:EFMu} of the $i^{th}$ throat, the inter-throat coupling and the shocks in \eqref{Eq:Model-Shocked} invoke a specific form of $T_{(i)\mu\nu}$ composed of ingoing or outgoing null matter:
  \be\label{Eq:nullemt}
 T_{(i)uu}(r,u) = f_i(u), \quad T_{(i)ur}(r,u) = T_{(i)rr}(r,u) = 0.
 \ee
 with
 \be\label{Eq:nullemt1}
 f_i(u) = - \la\(\bQ_{i-1}+ \bQ_{i+1}-2 \bQ_i\)\dt \bq_i' + \sum_A e_{i,A} \delta(u - u_{i,A}),
 \ee
 the right hand side of the first set of equations in \eqref{Eq:Model-Shocked}. It is easy to check that the em-tensor \eqref{Eq:nullemt} is conserved locally in the background metric \eqref{Eq:EFMu}. This em-tensor has a piece that involves dilute (continuous) infalling/outgoing energy due to inter-throat coupling and the pieces proportional to delta functions involving energy injection via the shocks which infall along null geodesics $u = u_{i,A}$. The dilaton takes a remarkably simple form 
 \be
 \Phi_i(r,u) = 2/r
 \ee
 in each throat as explicitly shown in Appendix \ref{App:Dilaton}. One may readily note from our previous discussion that in the microstate solutions with or without hair charge oscillations $T_{(i)\mu\nu} = 0$ because the right hand side of \eqref{Eq:Model-Shocked} vanishes in these solutions. These get turned on only after the injection of the first shock. 
 
 Clearly at the moment of injection, $u_{i,A}$, the mass of the corresponding black hole jumps by
 \be\label{Eq:M-jump}
 \delta M_i(u_{i,A}) = e_{i,A}
 \ee
 while the $SL(2,R)$ charges jump by
 \be\label{Eq:Q-jump}
  \delta \Q^0_i(u_{i,A}) =  -\frac{1}{2\tau_i'(u_{i,A})}e_{i,A}, \quad \delta \Q^\pm_i(u_{i,A}) = - \frac{e^{\pm \tau_i(u_{i,A})}}{2\tau_i'(u_{i,A})}e_{i,A}
 \ee
 as evident from the identities \eqref{Eq:Q-ids} where we have set $\beta = 2\pi$ (recall $M_i = - 2 \,{\rm Sch_i}$). Note $\delta \bQ_i \dt \delta \bQ_i = 0$ as $\delta \bQ_i$ is the $SL(2,R)$ charge vector of an ingoing null shock. It is easy to see from \eqref{Eq:tau-p} and \eqref{Eq:tau-pp-ppp} that the above discontinuities in the $SL(2,R)$ charges imply that $\tau'_i(u)$ and $\tau''_i(u)$ remain continuous at $u = u_{i,A}$ while $\tau'''_i(u)$ is generically discontinuous then. This is consistent with \eqref{Eq:Model-Shocked} being fourth order in $\tau_i(u_i)$. Furthermore, using \eqref{Eq:tau-p} we readily see that for \eqref{Eq:M-jump} and \eqref{Eq:Q-jump} we get
 \be
 \delta M_i = 2\bQ_i \dt \delta \bQ_i
 \ee
 as should follow by varying $M_i = \bQ_i\dt\bQ_i$ and using $\delta \bQ_i \dt \delta \bQ_i = 0$. Finally it is also easy to see that if $\bQ_i$ satisfy the restrictions given by \eqref{Eq:1poss} then so will  $\bQ_i +\delta\bQ_i$ (since $e^x + e^{-x} \geq 2$ for real $x$), and therefore can be parametrized using \eqref{Eq:Q-par} in terms of $M_i$ satisfying $0 \leq M_i \leq Q$ and $\rho_i$ restricted to the range prescribed in \eqref{Eq:rho-range}. 
 
Before the shock, we assume that the system is in a typical black hole microstate described in the previous section. The full configuration is macroscopically characterized by the total mass $M$ and $Q$. The $AdS_2$ black holes have time-independent masses $0 \leq M_i \leq Q$ such that they sum to $M$, i.e. $\sum_i M_i = M = \me_{\Q}$. The $SL(2,R)$ charges at each lattice site is parametrized by \eqref{Eq:Q-par} and \eqref{Eq:rho-range} in terms of $M$, $M_i$, $Q$ and $\rho_i$. Futhermore, the hair charges $\bq_i$ assume the configuration \eqref{Eq:qmicro} which is a linear superposition of the static component $\bq_i^{loc}$ in \eqref{Eq:qloc} locked to the $\bQ_i$ (we choose the additional parameter $\boldsymbol{\mathcal{K}}$ randomly -- it has no bearing on the dynamics), and the monopole term $\bq_i^{mon}$ in \eqref{Eq:qmon} parametrized by $\alpha$. To choose a typical microstate for fixed $M$ and $Q$, we will need to randomly allocate $M_i$ and $\rho_i$ to the lattice sites subject to the restrictions mentioned.\footnote{If we choose initial conditions for $\tau_i$ using \eqref{Eq:init-2}, then we have to randomly choose $\tau_{i0}$ for the initial microstate also.} Furthermore, we will need to specify $\alpha$. 

We consider the initial microstate to be hairless, i.e. with $\bq^{rad}= 0$ with the expectation that these hair charge oscillations will decay via their coupling to the asymptotic region of the geometry. The phenomenological features of response to shocks are unaffected anyway in the presence of hair in the initial microstate. The Hayden-Preskill protocol however will become more complicated -- we will return to this issue later.

It is evident from \eqref{Eq:Model-Shocked} that 
\be
\sum_i \bq_i'' = 0
\ee
 implying that \be \sum_i {\q^0_i}' = \alpha. \ee
This time-independent monopole parameter $\alpha$ remains unaltered even when shocks are injected into the system. It is determined by the formation of the black hole. It turns out that we obtain the desired phenomenological features mentioned at the beginning of this section provided $\alpha \geq 0$ when $\lambda > 0$. We also recall from the discussion in the previous section that we must set $\lambda >0$ in order that the average energy of a microstate in the ensemble with fixed $M$ and $Q$ is positive.
 
 We will restrict ourselves here to a one-dimensional lattice (chain) with periodic boundary conditions. We can readily simulate  our model following the method of \cite{Joshi:2019wgi}. Noting that the first set of equations in \eqref{Eq:Model-Shocked} are actually fourth order equations for $\tau_i(u)$ (i.e. $t_i(u)$) we rewrite these as four first order equations for $\bQ_i(u)$ and $\tau_i(u)$ as below using \eqref{Eq:Q-ids} and \eqref{Eq:tau-p} (we set $\beta = 2\pi$):
 \ba\label{Eq:Model-Shocked-2}
{\Q^0_i}' &= &\frac{1}{2\tau_i'}\( \la\(\bQ_{i-1}+ \bQ_{i+1}-2 \bQ_i\)\dt \bq_i' - e_i \delta(u - u_i)\),\nonumber\\
{\Q^+_i}' &= &\frac{e^{\tau_i}}{2\tau_i'}\( \la\(\bQ_{i-1}+ \bQ_{i+1}-2 \bQ_i\)\dt \bq_i' - e_i \delta(u - u_i)\),\nonumber\\
{\Q^-_i}' &= &\frac{e^{-\tau_i}}{2\tau_i'}\( \la\(\bQ_{i-1}+ \bQ_{i+1}-2 \bQ_i\)\dt \bq_i' - e_i \delta(u - u_i)\),\nonumber\\
\tau_i' &=& \frac{1}{2}\(\Q^-_i e^{\tau_i}+\Q^+_i e^{-\tau_i}-2 \Q^0_i\).
\ea
With initial values for $\bQ_i$ chosen randomly (thus picking a typical microstate) and initial value of $\tau_i$ set by \eqref{Eq:init} or \eqref{Eq:init-2} as discussed before, we can readily simulate these equations together with the second set of equations in \eqref{Eq:Model-Shocked}. The initial values of $\bq_i$ and $\bq_i'$ are also chosen randomly such that the initial conditions describe a typical initial hairless microstate solution (with $\bq^{rad} = 0$) as discussed above. Note we can compute $\tau_i''$ and $\tau_i'''$ using \eqref{Eq:tau-pp-ppp} from $\bQ_i$ and $\tau_i$ also at any given instant.

\begin{figure}[h]
\includegraphics[width = \linewidth]{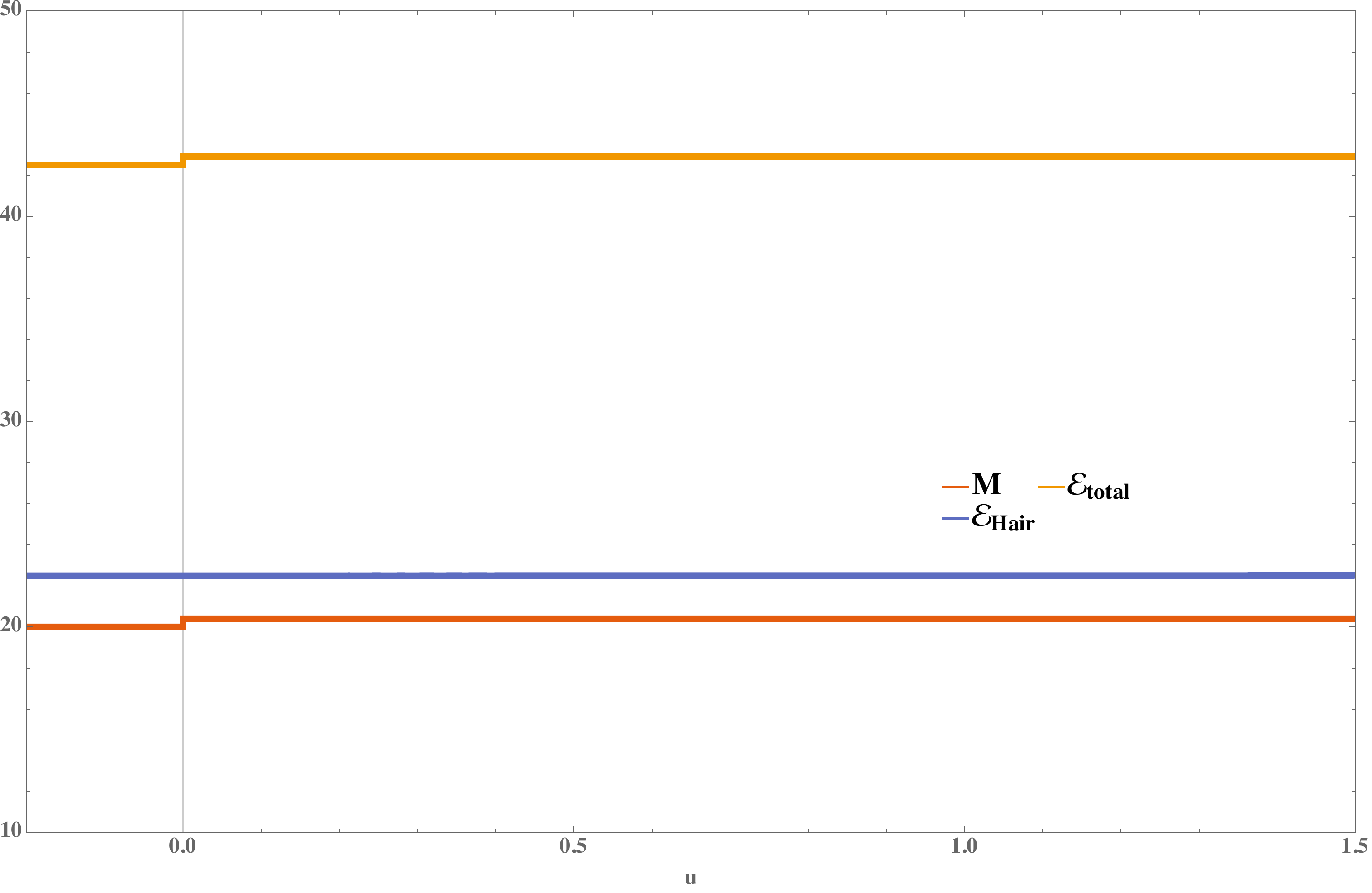}\caption{Evolution of $\me_{\Q}$ and $\me_{\q}$ after a shock in a lattice with 5 sites ($AdS_2$ black holes). Here $\la = 1$ and $\si = 0.01$. The shock  with $e = 0.4$ has been injected into the first site at $u= 0$. The variables $M_i$ and  $\bQ_i$  of the initial microstate solution have been chosen randomly as described in the text with the total initial mass $M= 20$ and $Q \approx 5.05$. Also $\alpha = 1$. Note that $\me_{\Q} = M$, the sum of the ADM masses of the $AdS_2$ throats, and $\me_{\q}$, the energy in the hair charges, are separately conserved (except at the moment of shock injection) to a very good approximation. Also approximately all the energy injected in the shock goes to $\me_{\Q}$. }\label{Fig:E-evolution-1-shock}
\end{figure}
 
\begin{figure}
\begin{center}
\begin{center}
\subfloat[Evolution of $\Q_i^0$ after the shock. 
\label{Fig:Q0-evolution-1-shock}]
{ \includegraphics[width=.5\linewidth]{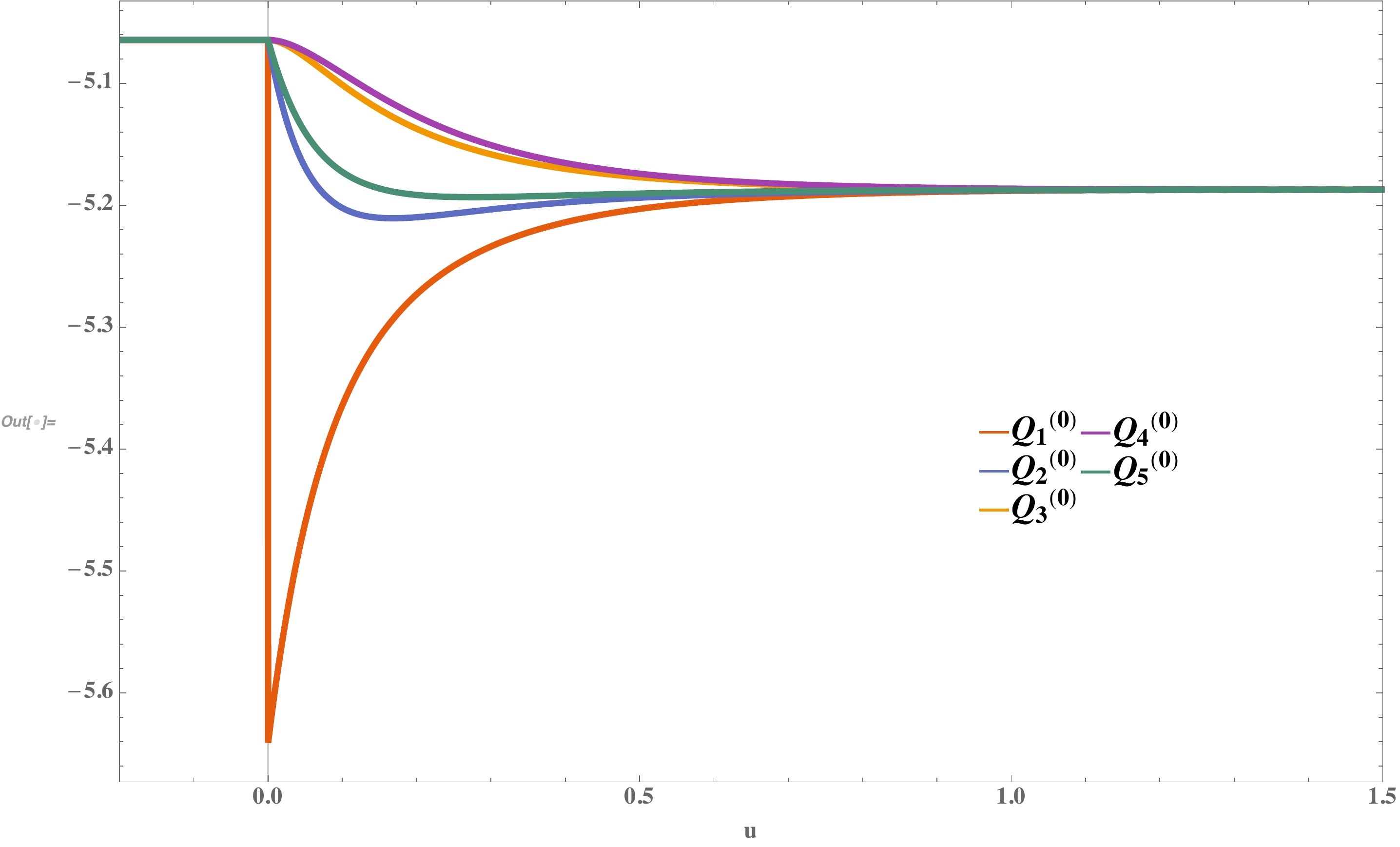}}
\subfloat[Evolution of $\Q_i^+$ after the shock.\label{Fig:Qp-evolution-1-shock}]
{ \includegraphics[width=0.5\linewidth]{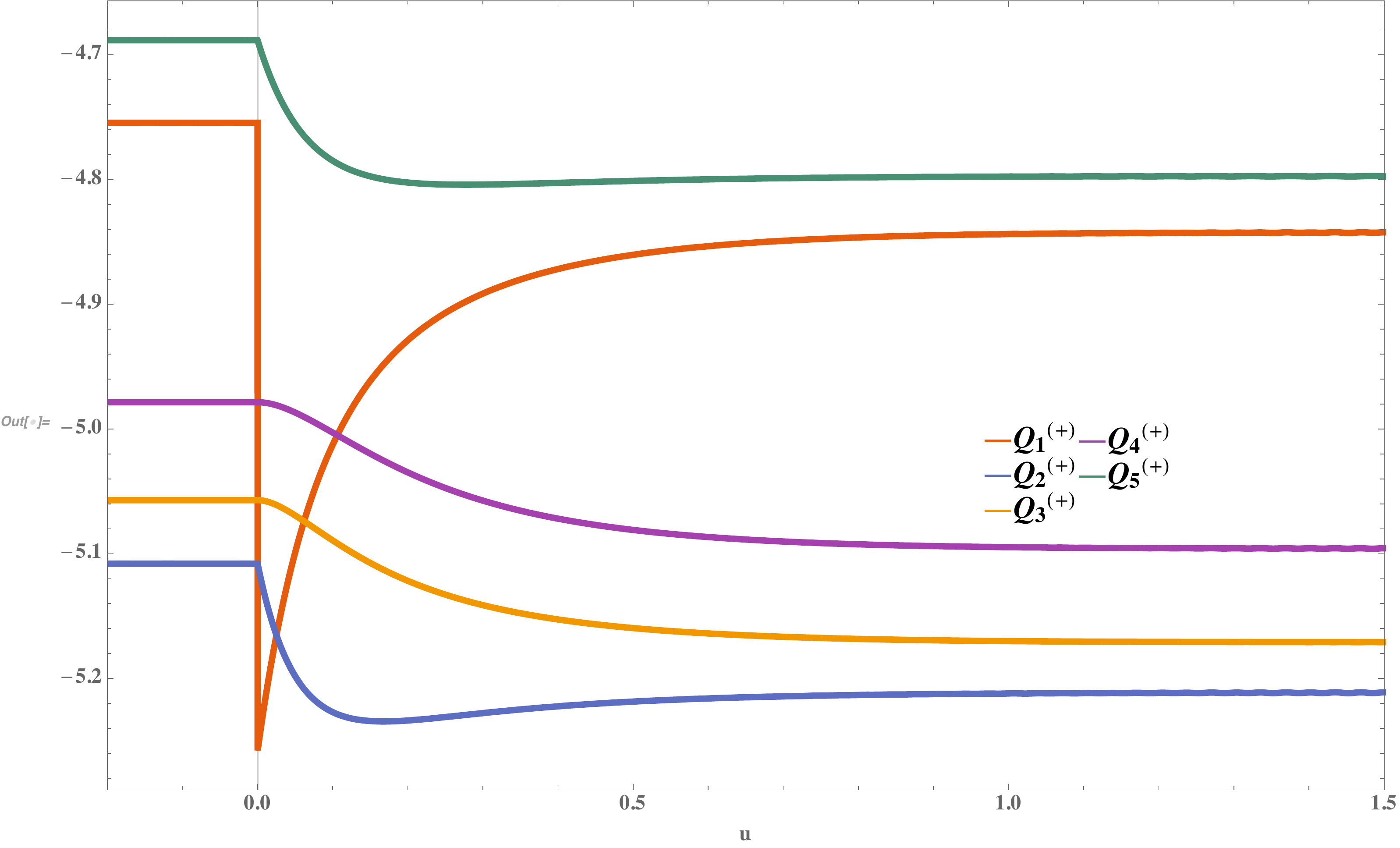}}
\noindent \\
\subfloat[Evolution of $\Q_i^-$ after the shock.\label{Fig:Qm-evolution-1-shock}]
{ \includegraphics[width=0.5\linewidth]{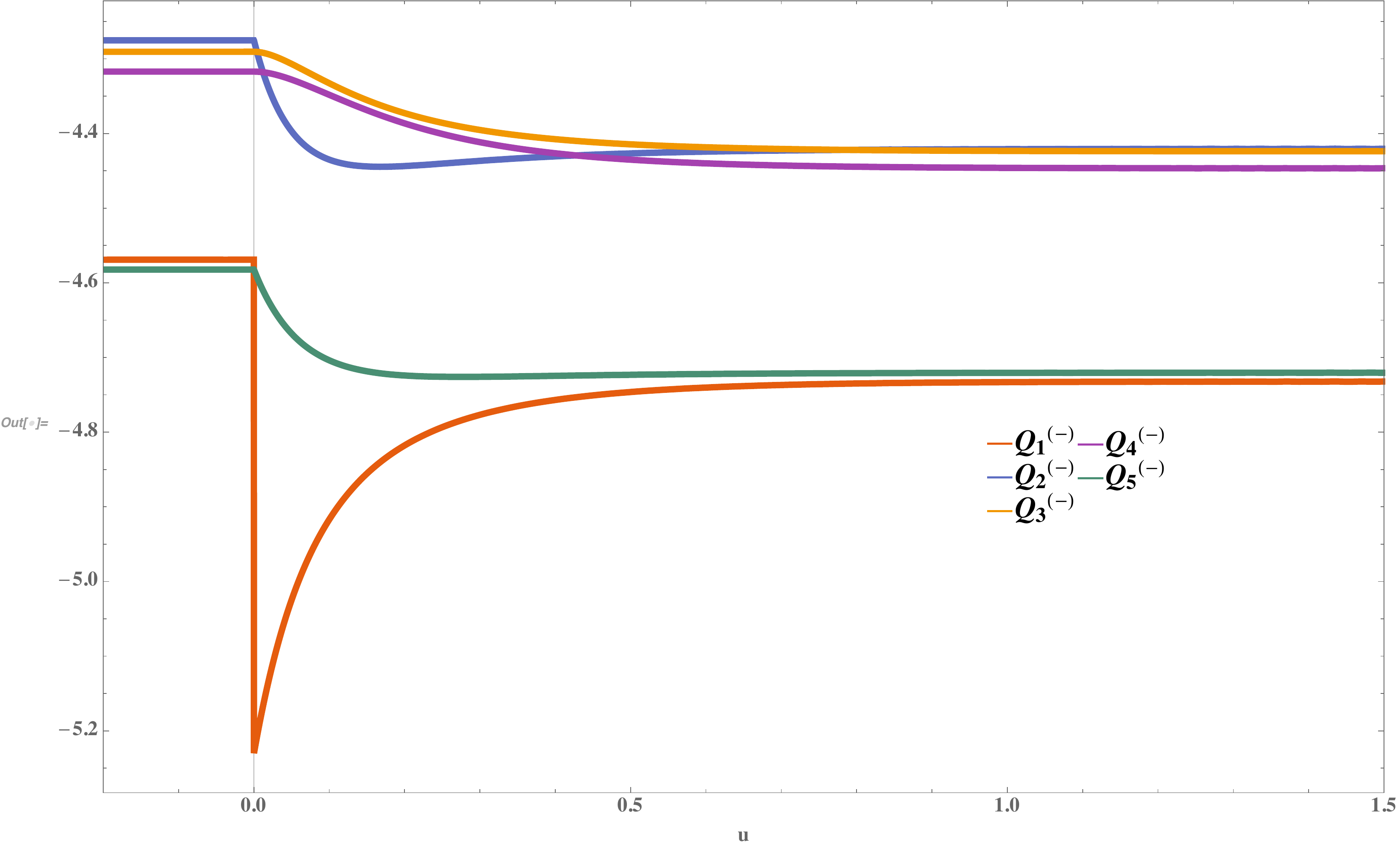}} 
\subfloat[Evolution of $M_i$ after the shock.\label{Fig:M-evolution-1-shock}]
{ \includegraphics[width=0.5\linewidth]{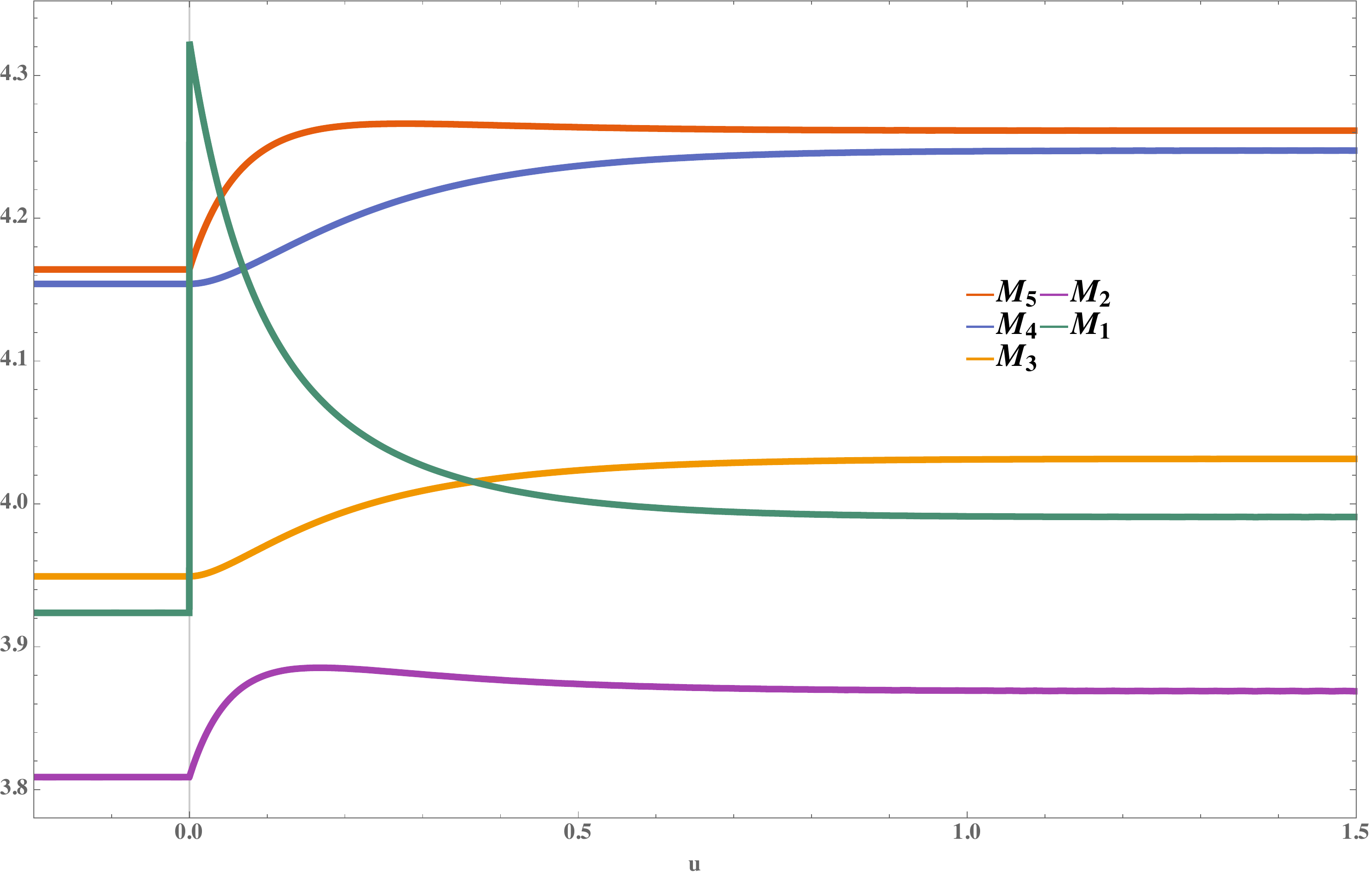}}
\end{center}
\caption{The evolution of the $SL(2,R)$ charges $\bQ_i$ and $M_i$ in our 5-site model after the first site is shocked at $u=0$. Here $\lambda$, $\sigma$, $\alpha$ and $e$ (the shock energy) are set as in Fig. \ref{Fig:E-evolution-1-shock}. We have chosen the initial conditions randomly corresponding to a typical initial microstate as exhibited above. The corresponding evolution of the energies $\me_{\Q}$ and $\me_{\q}$ are as in  Fig. \ref{Fig:E-evolution-1-shock}. It is clear from the above plots that all  $M_i$ and thus $\bQ_i$ relax eventually to constant values with $\bQ^0_i$ converging to the same value at each lattice site. This implies that the full system relaxes to a final microstate solution.}\label{Fig:Q-evolution-1-shock} 
\end{center}
\end{figure}

 We report our results for a 5 site chain with periodic boundary conditions which hold also for smaller or larger number of sites. We choose initial conditions describing a typical initial microstate solution and set $\la = 1$ and $\si = 0.01$. The phenomenology does not depend on these choices provided $\alpha \geq 0$. We can readily observe from the right hand side of the first equation in \eqref{Eq:Model-Shocked} that $\lambda\alpha$ acts like a diffusion constant which needs to be non-negative. Since $\lambda$ is positive, $\alpha$ has to be non-negative. The features listed below hold even if the microstate that we shock is hairy. 
\begin{enumerate}
\item We note that even after the shock,  the total mass $M := \me_{\Q}$ and the energy in the hair quanta $\me_{\q}$ are separately conserved to a very good degree of approximation as shown in Fig. \ref{Fig:E-evolution-1-shock} although the full system has a very complicated time-evolution (see Fig. \ref{Fig:Q-evolution-1-shock}). Although $\me_{\Q}$ and $\me_{\q}$ look constant on the scale of Fig. \ref{Fig:E-evolution-1-shock}, actually there are tiny fluctuations of $\mathcal{O}(10^{-4})$ which eventually decay away. Note at each site there is exchange of energy between the two sectors -- each $M_i$ vary significantly with time as shown in Fig. \ref{Fig:M-evolution-1-shock}. It is only the sum of the ADM masses given by $\me_{\Q}= M$ which remains almost approximately unperturbed. We have checked that if we increase the number of lattice sites ($n$) and take the continuum limit keeping mass and charge densities finite, then the ratio of the amplitude of these fluctuations $\Delta M$ to the total mass existing in the transition time go to zero, i.e. $\Delta M/M \rightarrow 0$, in the continuum limit.  Therefore, the separate conservation of $\me_{\Q}$ and $\me_{\q}$ becomes an increasingly better approximation as we approach the continuum limit.
\item It is also clear from Fig. \ref{Fig:E-evolution-1-shock} that the energy of the shock is almost fully absorbed by $\me_{\Q}$ , the sum of the ADM masses of the black hole. The ratio of the increase in the total mass to the energy in the injected shock approaches unity in the continuum limit.

\item Finally, it is clear from Fig. \ref{Fig:Q-evolution-1-shock} that the full system relaxes to another microstate solution since all $M_i$ (and therefore $\Q_i$) relax to constant values. Particularly, as evident from Fig. \ref{Fig:Q0-evolution-1-shock}, $\Q^0_i$ converge to the same value at each lattice site giving the value of $Q$ for the final microstate that is different from the corresponding initial value.\footnote{The $SL(2,R)$ frame given by the direction along which $\bQ_i$ takes the same value in the final microstate solution cannot change as it is determined by the monopole component of $\bq_i'$ which as noted above is a constant of motion with magnitude $\alpha$.} The final microstate also supports coherent hair oscillations $\bq_i^{rad}$ along with $\bq_i^{loc}$ locked with the final $SL(2,R)$ charges as in \eqref{Eq:qloc} and the monopole term $\bq_i^{mon}$ with the same $\alpha$ as in \eqref{Eq:qmon}. The hair oscillations $\bq_i^{rad}$ are freely propapagating, i.e. satisfy \eqref{Eq:qrad}, \eqref{Eq:KGj} and \eqref{Eq:qrad2}. They are thus decoupled from the final microstate solution background. Note the full energy $\me_\q$ in the hair charges in the final microstate solution is then composed of three pieces as shown in \eqref{Eq:Eq-split-2} which include the contribution $\me_\q^{rad}$ from the coherent inhomogenous oscillating terms.
\end{enumerate}

Thus we find that our model is phenomenologically viable with the monopole parameter $\alpha >0$. It has the crucial property of relaxation like a black hole, i.e. an initial black hole microstate evolves to a final black hole microstate after a perturbation by a shock with the energy in the shock absorbed almost entirely by $M$, the sum of the ADM masses of the $AdS_2$ throats. This also holds if we also perform multiple shocks. Furthermore, $M$ remains approximately constant even during the transitory period between two microstates except for the moment of shock injections. These features become exact in the continuum limit if we consider appropriate ratios as mentioned above. 

It is easy to see that the approximate conservation of $\me_\Q$ implies that if we sum over all the throats at any instant $u$, then
\begin{align}\label{Eq:nullemt3}
 \sum_iT_{(i)uu}(r,u) &\approx \sum_A e_{i,A} \delta(u - u_{i,A}),\nonumber\\
  \quad \sum_i T_{(i)ur}(r,u) &= \sum_i T_{(i)rr}(r,u) = 0.
 \end{align}
The above is also borne out in our numerical solutions to an excellent approximation even in presence of multiple shocks. Therefore, the total inflow/outflow of energy in the throats due to inter-throat coupling alone should approximately vanish at intermediate times. (Of course, for both initial and final microstate solutions the inflow/outflow of energy vanishes exactly for each throat individually).

We note that our claim is that these phenomenological features hold for a typical (randomly chosen) initial microstate solution even if it supports hair. However, it is possible that these features actually hold for all microstate solutions since we have not encountered any exception so far. 

It is important to note that our model has inherent pseudorandomness. Given an initial state and perturbations in the form of a few shocks, it is quite hard to determine analytically the details of the final microstate, except for the total mass $M$ approximately (since it is approximately the sum of the initial total mass and the energy injected by the shocks) and the monopole parameter $\alpha$ which is a constant of motion. Even the final value of the macroscopic order parameter $Q$ is hard to predict as it depends on the details of the initial microstate, i.e. not on the initial values of $Q$ and $M$ and $\alpha$ alone. The value of $Q$ in the final microstate varies by at least 1 percent for different choices of initial microstates (with same values of $Q$ and $M$ and $\alpha$). The Hawking radiation emitted from the final microstate can be expected to inherit the pseudorandomness of the background semiclassical black hole dynamics. 

One can also explore the quantum chaotic features in our model via the study of the differences in $\tau_{i0}$ parameters in $\tau_i(u)$ between the initial and final microstates. We leave it to a future study as the out-of-equilibrium OTOCs merit a separate study on their own right.

\begin{figure}
\begin{center}
\begin{center}
\subfloat[The relaxation time as a function of the mass density when $Q-$density is 1 and $\alpha = 2$.\label{Fig:SatTimeM}]
{ \includegraphics[width=\linewidth]{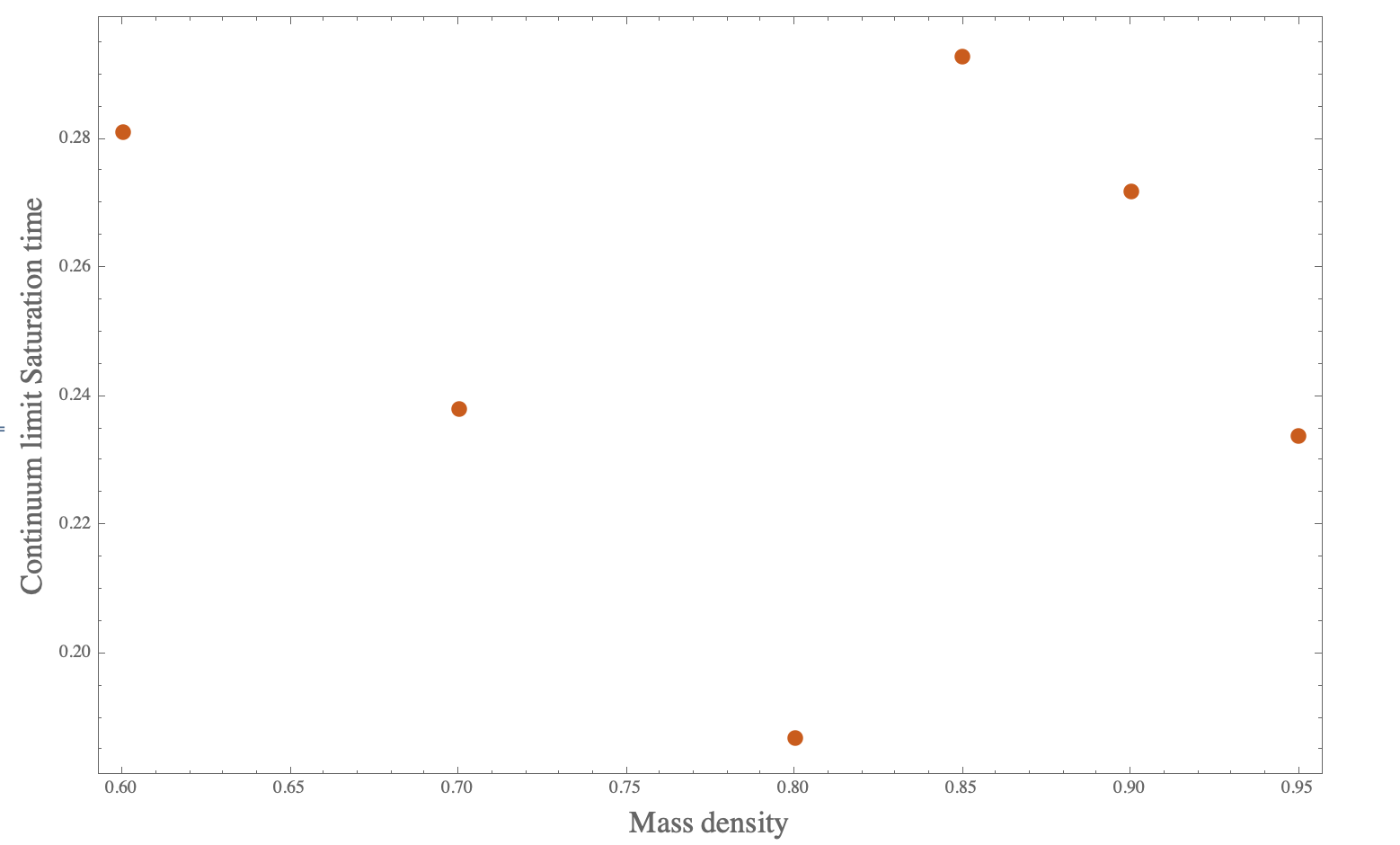}}\\
\subfloat[The relaxation time as a function of $Q$ when $M-$density is 0.8 and $\alpha = 1$ in a 5-site model. The fitted line is $0.004 +0.365 e^{-0.624 Q}$.\label{Fig:SatTimeQ0}]
{ \includegraphics[width=\linewidth]{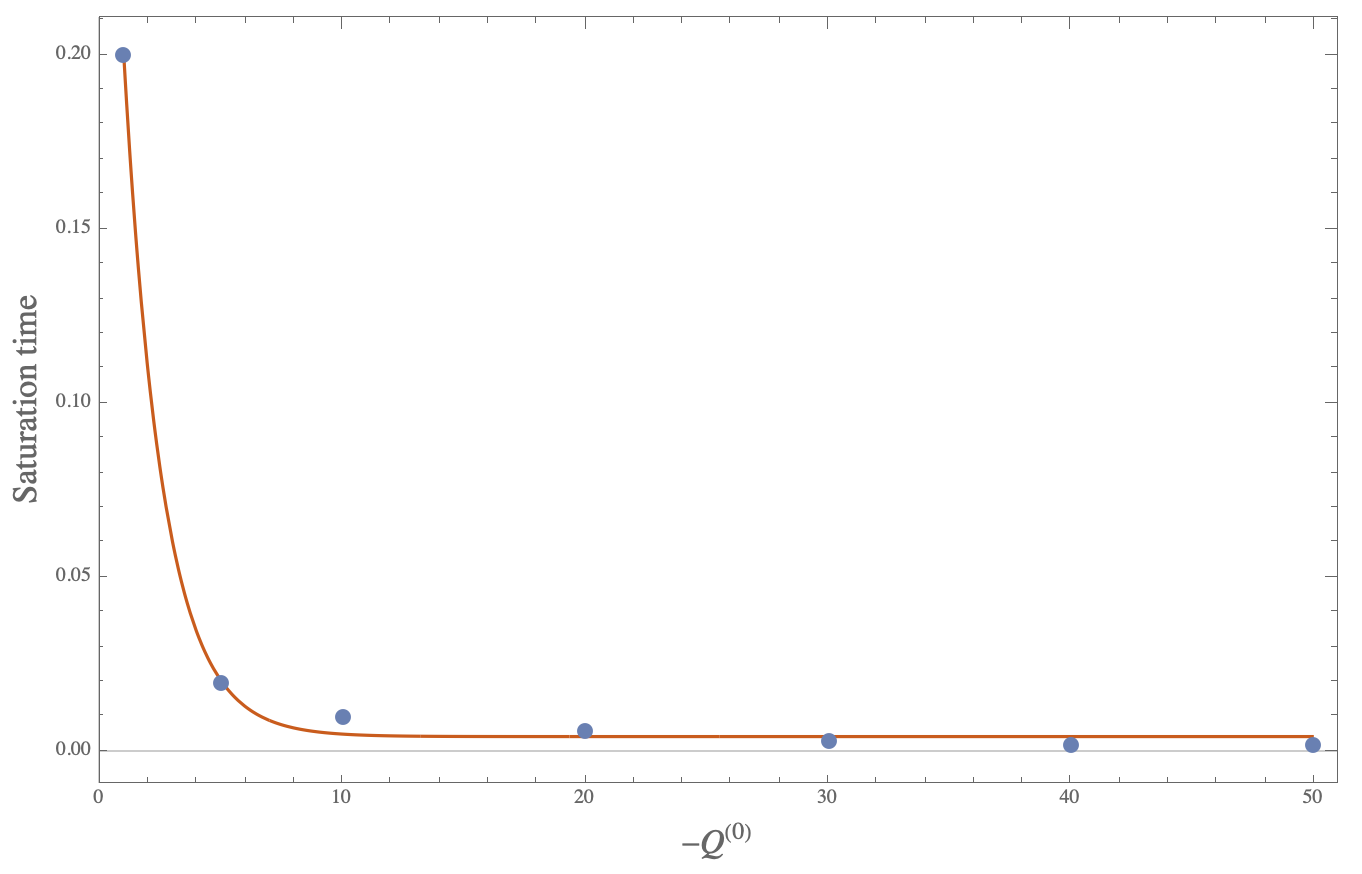}}
\end{center}
\caption{The dependence of the relaxation time on $M$ and $Q$.  Note there is about 1-2 percent variation over choice of initial microstates. This is not shown above.}\label{Fig:Saturation-time}
\end{center}
\end{figure}
The transit time between microstates depends on the macroscopic parameters $Q$, $M$ and $\alpha$ of the initial macrostate, and not significantly on the energy injected by the shock. We define the relaxation time as the time when the $SL(2,R)$ charges attain $99$ percent of their final values. Note that only the combination $\lambda \alpha$ is significant for the dynamics so we can set $\lambda = 1$. It also turns out that $\sigma$ does not influence the relaxation time significantly.

In Fig. \ref{Fig:SatTimeM}, we have studied the relaxation time in the continuum limit for different mass densities and with fixed unit $Q-$density and $\alpha = 2$. There are statistical fluctuations of about 1-2 percent. The relaxation time does not vary much over the range of mass densities studied.  We find that if we increase $Q$ keeping $M$ and $\alpha$ fixed, then the relaxation time decreases rapidly and goes to $0.004$ for large $Q$ density when mass density is $0.8$ and $\alpha = 2$. In Fig. \ref{Fig:SatTimeQ0}, we present the dependence of the relaxation time on $Q$ for fixed $M$ and $\alpha$. We conclude that the relaxation time is small compared to the mass density if we allow $Q$ to vary over the ensemble of microstates with fixed mass density and $\alpha$.\footnote{We choose $\alpha$ by assuming equipartitioning of energy between the black hole masses and the monopole radiation via quantum fluctuations.} In the next section, we will identify the relaxation time with the Hayden-Preskill time by showing that the information of the infalling shocks is encoded by the black hole dynamics in the coherent hair oscillations $\bq_i^{rad}$ which decouples from the final microstate to which the full system relaxes.

 \section{The Hayden-Preskill protocol}\label{Sec:HP}
 Suppose Alice wants to erase some top secret information. She thinks that the best way to do this would be to throw it into a black hole.\footnote{The Landauer principle \cite{BENNETT2003501} suggests that it is a good idea since the entropy of the black hole increases in this procedure.} Alice would like to know if she throws the information into our model black hole, then whether Bob, who is outside the black hole, can decode the message after some time has passed or not. This is exactly the  question that we will answer in this section. 
 
\begin{figure}[t]
\includegraphics[width=\linewidth]{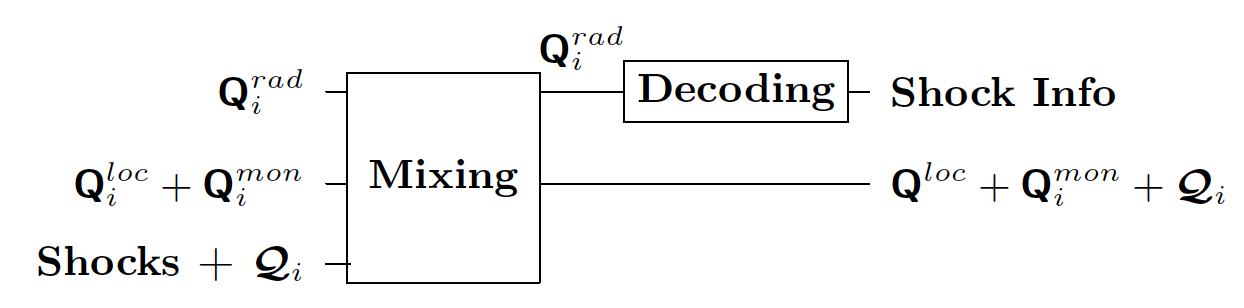}
\caption{\textit{Mixing} indicates the dynamics of our model which mixes up all the degrees of freedom with the external shocks and \textit{Decoding} is the procedure that Bob uses to extract information about the shocks. We will consider the situation where the hair oscillations ($\bq^{rad}_i$) are initially zero to simplify our decoding protocol. After the model relaxes to a new microstate with some hair oscillations on top, Bob applies his decoding procedure on the decoupled hair $\bq^{rad}$ and recovers some classical information about the shocks, namely their positions and time ordering. }
\label{Fig:HaydenPreskill}
\end{figure}

We assume that Alice's message is in the form of one or two compact classical bits going into one or two of our (nearly) $AdS_{2}$ throats. We restrict ourselves to studying the decoding of the classical information encoded in these shocks via their positions and time ordering. We ask a very concrete question: can Bob eventually decode the locations and time ordering of these shocks which carry Alice's message? We have seen in Sec. \ref{Sec:Shock} that after all the shocks have gone in, the system relaxes to another microstate solution with decoupled hair oscillations on top.  In what follows, we will show that the black hole dynamics encodes the classical information of the positions and time-ordering of the shocks in the hair charge oscillations $\bq_i^{rad}$ that decouple from the final microstate solution to which the system relaxes. Although the information is classical, the quantum-circuit like diagram in Fig.~\ref{Fig:HaydenPreskill} is a good schematic representation of the decoding process. We expect that this information will be further encoded into the gravitational/Hawking radiation which interacts with $\bq_i^{rad}$. However, Bob can choose to decode the information from $\bq_i^{rad}$ directly by probing these gapless degrees of freedom that live on the surface of the horizon.
 
 To simplify the decoding procedure, we will assume that the initial black hole microstate is not hairy. This assumption is justified on the grounds that the hair charge oscillations $\bq_i^{rad}$ should decay by virtue of their coupling to the asymptotic region of the geometry in the form of gravitational radiation, etc. Otherwise, we should think of the preexisting hair oscillations as \textit{early radiation} (not to be confused with early Hawking radiation) and the decoding procedure, which has to extract information for the full $\bQ_i^{rad}$ that decouples from the final microstate, will depend on the details of this \textit{early radiation} which should be known to the decoder a priori. We will not consider this complexity here since we have physical mechanisms of hair removal anyway. 
 
\subsection{Decoding a single shock}
\label{Subsection:SingleShockDecode}
\begin{figure}[t]
	\includegraphics[width=\linewidth]{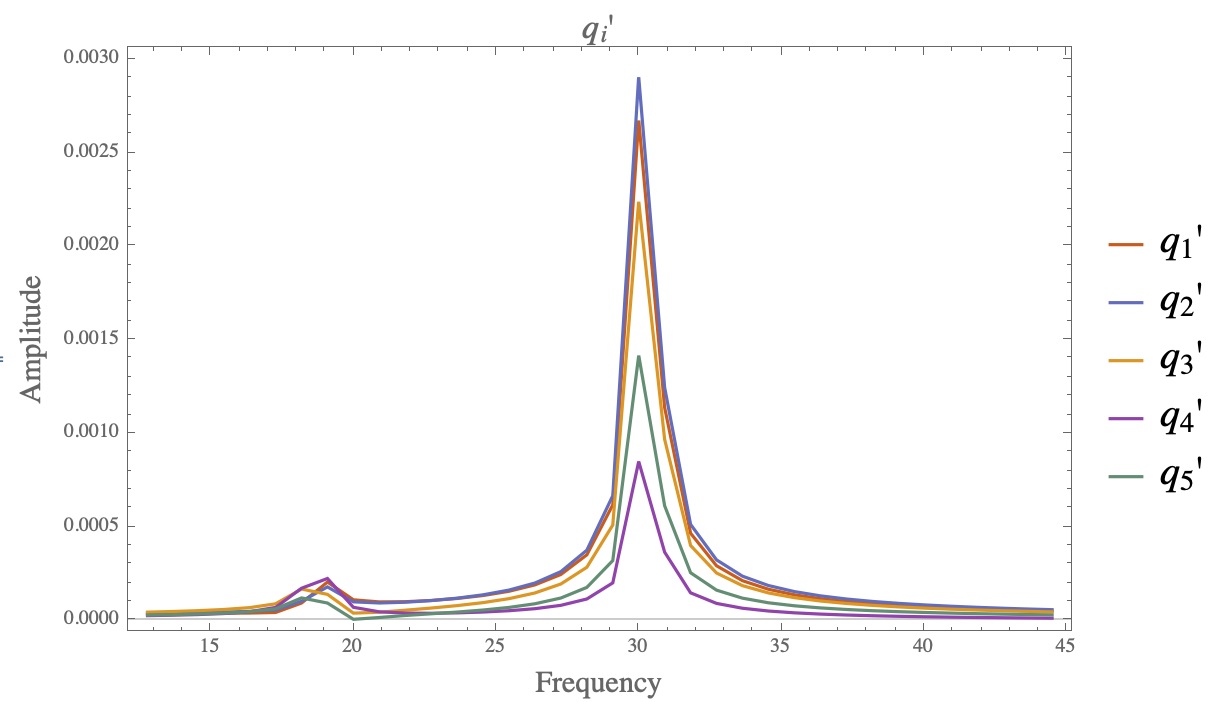}\caption{Amplitudes of the Fourier transform of $q_i'$ for each site obtained after the full system relaxes to the final microstate. The peaks coincide with the numerically computed normal mode frequencies to a very good approximation. The finite width of these peaks is due to finiteness of the time-window and also numerical noise.}\label{Fig:Fourier}
\end{figure}

We assume that Alice shocks only one site in a 5 site chain and ask what Bob must do in order to recover the position of the single shock from the hair oscillations. We recall from \eqref{Eq:qrad} that the hair charge oscillations $\bQ_i^{rad}$ are parametrized by $q_i$ and these follow the free lattice Klein-Gordon equation \eqref{Eq:KGj}. For the case of a 5 site lattice, we can readily obtain the normal mode frequencies for these coupled oscillators $q_i$ (that are decoupled from the lattice charges). With the same parameters as those chosen in the context of Fig.~\ref{Fig:E-evolution-1-shock} and \ref{Fig:Q-evolution-1-shock}, the normal mode frequencies turn out to be $(\pm 30.2731, \pm18.7098,0)$. Each of these frequencies are doubly degenerate so that we have 10 normal modes. The zero frequency modes are monopoles and cannot appear in the final ${q_i}'$ due to conservation of the monopole term (see the discussion in the previous section). We perform a Fourier transform of the numerical data for $q_{i}^{\prime}$ for each site after the system relaxes to the final microstate and plot the amplitudes in Fig.~\ref{Fig:Fourier}. We only plot the positive frequencies because the amplitudes of the corresponding negative frequencies have to be the same (and the phases should be just reversed in sign). From the Fourier transformed data, we observe that peaks coincide with the analytically computed frequencies to a very good approximation.

 Remarkably, Bob can extract the position of the shock by looking at the phase differences between these two frequency modes at each site. The phase differences when the first site is shocked are found in Table \ref{tab:DecodingSingle}. We define the phases modulo $2\pi$.

\begin{table}[ht]
\centering
\begin{tabular}[t]{ cc } 
\hline
Site & Phase difference\\
 \hline
 1& 4.159\\
 2& 3.187\\
 3& 2.356\\
 4& 2.347\\
 5& 3.185\\
 \hline
\end{tabular}
\caption{Decoding a single shock at site 1}
\label{tab:DecodingSingle}
\end{table}

A clear pattern is visible. Sites 2 and 5 which are equidistant from the shocked site 1 have the same phase difference within our numerical accuracy. Similarly, sites 3 and 4 which are equidistant from the shocked site 1 also have nearly the same phase difference within our numerical accuracy. This symmetry clearly tells Bob that site 1 was shocked. This symmetric pattern repeats when any other site is shocked and also for any randomly chosen initial microstate. It generalizes as well if we have higher number of sites -- the symmetry in the site-dependent phase difference between the two largest positive frequency modes reveals which site has been shocked. 

This is a very non-trivial result because both the initial and final microstates are highly asymmetric. If some features of the decoupled hair oscillations are insensitive to the lattice charges but sensitive to the shock, then the decoding is possible. Otherwise the decoder needs to have access to the initial black hole microstate, which is forbidden in the Hayden-Preskill protocol. We conclude that the mentioned phase difference is a feature of the mentioned type that decodes the position of the shock. The energy of the shock, of course, can be obtained simply from the difference between the final and initial black hole masses to a very good approximation.

\subsection{Decoding two shocks}
\label{Subsection:DoubleShockDecode}
 If Alice throws in two shocks with equal energies at two distinct lattice sites in our model, Bob can decode the position and time ordering of the shocks by using a procedure similar to that described before. After the model relaxes to a new microstate, Bob should again study the Fourier transform of the decoupled radiation $q_{i}^{\prime}$. The pattern of phase differences will tell him where the shocks happened and also which shock happened first. For a 5 site chain there are exactly two possibilities, the shocks are either nearest neighbour or they are separated by one site. For example we consider two nearest neighbour shocks first at site 1 and then at 5. 
 The time difference between the two shocks is less than the time it takes for the model to relax to a microstate after a single shock. The phase differences are shown in Table~\ref{tab:Decoding15}.
 
 \begin{table}[ht]
\centering
\begin{tabular}[t]{ cc } 
\hline
Site & Phase difference\\
 \hline
 1& 1.599\\
 2& 2.747\\
 3& 3.783\\
 4& 4.844\\
 5& 5.994\\
 \hline
\end{tabular}
\caption{Decoding a shock at site 1 followed by a shock at site 5}
 \label{tab:Decoding15}
\end{table}

Keeping the shock size and time difference the same, we now reverse the time ordering of the shocks. That is, site 5 is shocked first followed by site 1. The phase differences are shown in Table~\ref{tab:Decoding51}.

\begin{table}[ht]
\centering
\begin{tabular}[t]{ cc } 
\hline
Site & Phase difference\\
 \hline
 1& 4.982\\
 2& 4.531\\
 3& 3.634\\
 4& 2.858\\
 5& 0.926\\
 \hline
\end{tabular}
\caption{Decoding a shock at site 5 followed by a shock at site 1}
\label{tab:Decoding51}
\end{table}

 A pattern can be easily noticed. In order to determine the position of the two shocks, Bob simply needs to determine the maxima and minima of the phase differences. The sites with the largest and smallest phase differences are the two sites that were shocked. If these two sites are nearest neighbour, then the site with the smallest phase difference was shocked first and the site with the largest phase difference was shocked later. This pattern holds for any randomly chosen initial microstate.
 
 The only other possibility for a 5 site model is when the shocks are separated by a single site. For example, we consider shocking first site 1 and then site 3, which is shown in Table~\ref{tab:Decoding13}. Similarly, the phase differences when site 3 is shocked first, followed by site 1 are shown in Table~\ref{tab:Decoding31}.
 
 \begin{table}[ht]
 \centering
\begin{tabular}[t]{ cc } 
\hline
Site & Phase difference\\
 \hline
 1& 5.837\\
 2& 3.639\\
 3& 1.986\\
 4& 4.852\\
 5& 2.709\\
 \hline
\end{tabular}
\caption{Decoding a shock at site 1 followed by a shock at site 3}
 \label{tab:Decoding13}
\end{table}

\begin{table}[ht]
\centering
\begin{tabular}[t]{ cc } 
\hline
Site & Phase difference\\
 \hline
 1& 2.374\\
 2& 3.553\\
 3& 6.166\\
 4& 2.824\\
 5& 4.591\\
 \hline
\end{tabular}
\caption{Decoding a shock at site 3 followed by a shock at site 1}
\label{tab:Decoding31}
\end{table}

In this case as well, the sites with maximum and minimum phase differences are the locations of the two shocks. However, when these locations are separated by one site, the site with the maximum phase difference was shocked first. This feature once again holds for an arbitrarily chosen initial microstate.

A natural question to ask next would be if the decoding procedure holds for two shocks with unequal shock energies. This is indeed the case if the differences between the shock energies is small ($ < 0.3$). The decoding procedure fails for larger differences in the shock energies. This could be a limitation of the classical approximation. In principle, we should treat the hair charge sector as an open quantum system interacting with the bath in the form of the lattice of $AdS_2$ throats. We leave this to a future study. 

\subsection{Algorithm}
\label{Subsection:Algorithm}
We can combine the discussion from the previous two subsections into a unified algorithm that Bob can utilize to extract information about the shock position(s) and time ordering. It is assumed that Alice only shocks either one or two sites in the 5 site chain. The decoding algorithm is then as follows:

\begin{enumerate}
	\item Compute the Fourier transform of the $q_i^{\prime}$ at each site.
	\item From the Fourier transform data, compute the phase difference between the two positive frequency modes at each site.
	\item If these phase differences are symmetric about a specific site, then this site has been shocked once or twice. 
	\item If the phase differences are not symmetric, then determine the maximum and minimum phase differences. These are the locations of the shocks. 
	\item If these two sites are nearest neighbours, then the site with the smallest phase difference was shocked first.
	\item If these two sites are separated by one site, then the site with the largest phase difference was shocked first.
\end{enumerate}
The above algorithm for decoding the locations and time orderings of the shocks is likely to generalize to the case of higher number of sites -- we will need to note the phase differences of the three or more positive frequency normal modes. Unfortunately, there are several numerical issues in performing the Fourier transforms of $q_i'$. For higher number of sites, the procedure for the noise correction discussed in Appendix \ref{App:Drift} to get reliable Fourier transforms becomes harder to implement. 
These issues have made it difficult for us to generalize our algorithm for 6 or more sites. In Appendix \ref{App:Decode34}, we discuss the case for lower number of sites.

In the case of multiple shocks with not large differences between their energies (and which all happen within the relaxation time of the system), we can also look into inter-site phase differences in the Fourier modes of the highest frequencies, etc. It is quite likely that if the number of shocks are not too many, a generalized decoding protocol for the locations and time-ordering of the shocks exists. However, we have not been able to find a simple decoding procedure to tell us if the same site has been shocked once or twice. Understanding the mechanism that makes our protocol work better can help us to address this issue. We also expect that one can setup a more efficient protocol with the quantized hair charges and its complexity will be commensurate with that of the information in the infalling bits and not of the black hole interior as in the algorithm discussed above.

We also need to address if one can construct the Hayden-Preskill protocol for infalling qubits instead of infalling classical bits. This can be studied by introducing infalling quantum degrees of freedom inside the $AdS_2$ throats. Together with the quantized hair charges, they will form an interacting open quantum system exchanging energy and $SL(2,R)$ charges with the classical geometries of the $AdS_2$ throat lattice. This setup merits a separate study.

It is interesting to point out that in \cite{Joshi:2019wgi} a semi-holographic model of an impurity interacting with a strongly coupled quantum dot was considered. The displacement of the impurity from the confining dot was coupled to a scalar operator in the localized theory whose dual description was JT gravity coupled to a scalar field.  It was found that if the total system energy was positive, the impurity attained a terminal velocity extracting energy from the quantum dot. At late time, the ADM mass of the $AdS_2$ throat decayed while the $SL(2,R)$ charges grew as $e^{a u}$, and the exponent $a$ was related to the initial impulse given to the impurity.  The impurity here was not infalling but still was an external impulse extracting work from the quantum dot. This indicates there is scope for asking how our model may process such external stimuli which instead of falling into the throats simply interact with them via appropriate couplings.

\section{A discussion on $SL(2,R)$ networks}\label{Sec:Network}
A crucial question to ask is whether the phenomenological properties of information processing could hold for a more general class of $SL(2,R)$ lattice models than the one so far considered. 
The crucial element of our model \eqref{Eq:Model} which leads to the Hayden-Preskill scenario of information mirroring is the coupling between the lattice charges and the gravitational hair. 
\begin{figure}
\includegraphics[width = \linewidth]{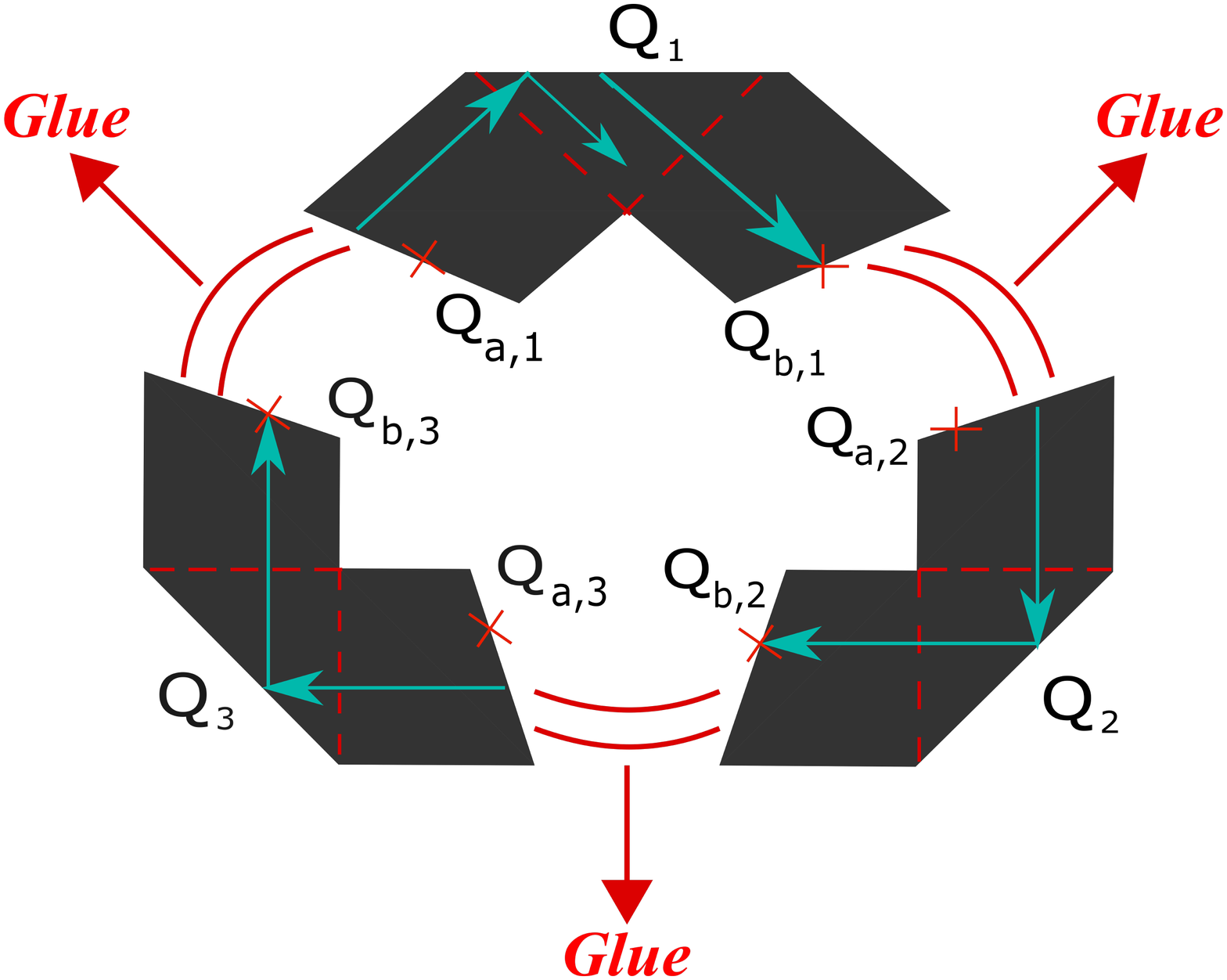}\caption{$AdS_2$ throats in a $3$ site chain can be networked via wormholes as depicted above. Each $AdS_2$ throat has a pant like structure bounded by two geodesics carrying charges $\bQ_{a,i}$ and $\bQ_{b,i}$ at $i^{th}$ site such that $\bQ_i = \bQ_{a,i} + \bQ_{b,i}$ and $\bQ_{a,i} = \bQ_{b,{i-1}}$ (also for $\bQ_{b,i} = \bQ_{a,{i+1}}$) for $i = 1,2,3$. The other edges are simply light-like geodesics -- the red dotted lines mark light-rays also. The pants are glued across the geodetic junctions with equal $SL(2,R)$ charges. It is clear that any light-like geodesic starting from a boundary can traverse the entire network -- note that the time-shifts while traversing the geodetic junctions -- the impact point at the other edge is marked with a cross. Here we show an example of the transmission of a shock through the network with perfectly transmitting conditions at the junctions and perfectly reflecting conditions at the boundaries.}\label{Fig:Network-AdS2}
\end{figure}

\begin{figure}
\includegraphics[width = \linewidth]{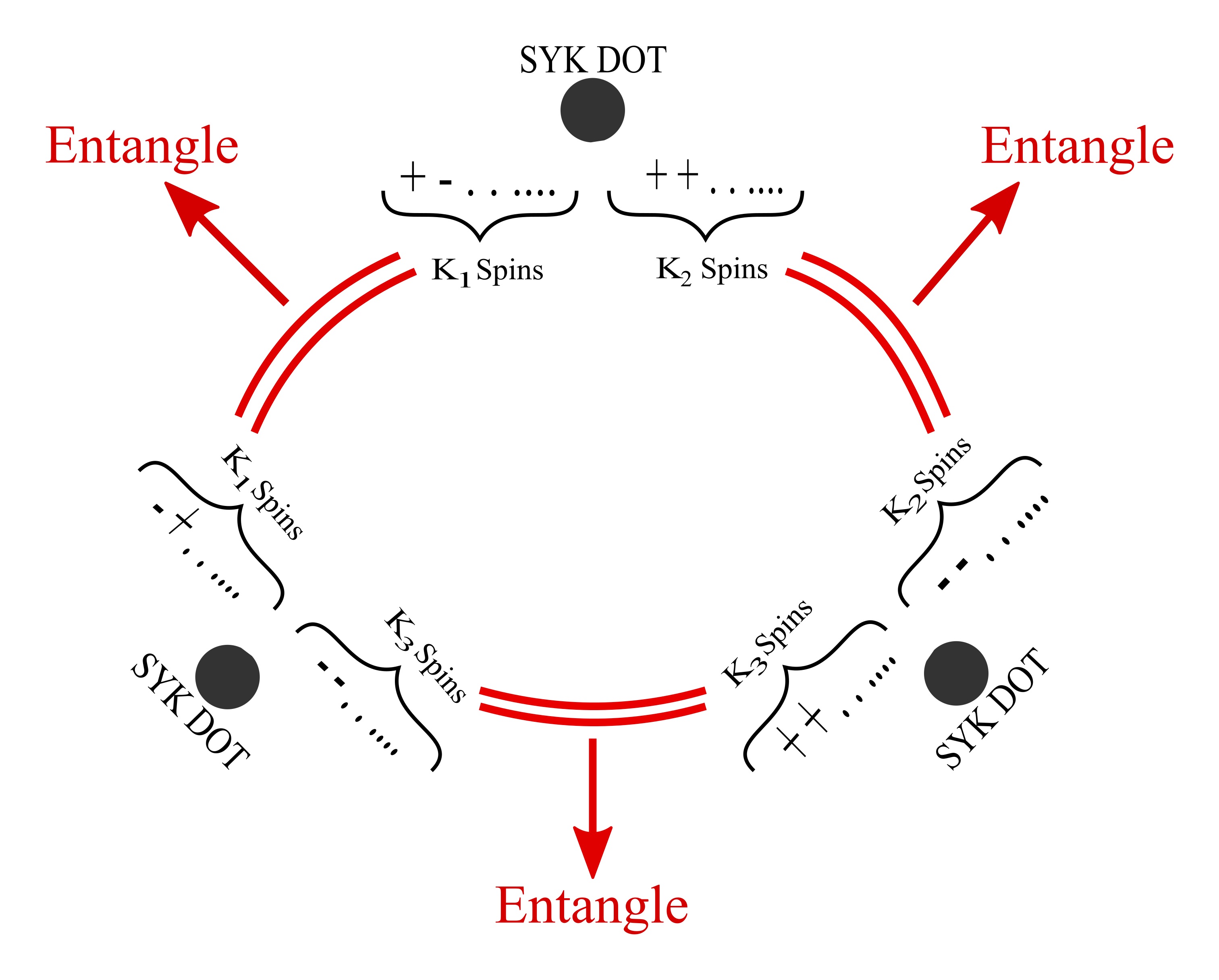}\caption{A tensor network of a chain of SYK spin states with periodic boundary conditions is illustrated above. We partition the $N_i$ spins into two blocks $N_{a.i}$ and $N_{b,i}$ such that $N_{a,i} = N_{b,{i-1}}$ (and thus $N_{b,i} = N_{a,{i+1}}$) for $i = 1,2,3$. We then maximally entangle the blocks of spins with the corresponding blocks of the neighboring sites as shown above. Finally, Euclidean evolution by the SYK Hamiltonians over each site (see text) symmatrizes over the $N_i$ spins at each site and different ways of pairing $2N_i$ Majorana fermions (with appropriate sign factors) at each site into $N_i$ pairs each of which is projected onto a spin state.}\label{Fig:Network-SYK}
\end{figure}

We can readily generalize our model such that the coupling between the two charge sectors is exactly as shown in \eqref{Eq:Model} and thus the total energy remains of the form \eqref{Eq:E} in the absence of perturbations. The only difference is that instead of coupling the $AdS_2$ throats only at their boundaries via nearest neighbor lattice couplings, we can also form Lorentzian wormhole networks as shown in Fig. \ref{Fig:Network-AdS2}. In the simplest version, each $AdS_2$ throat in a chain ends on two geodesics carrying $SL(2,R)$ charges $\bQ_{a,i}$ and $\bQ_{b,i}$ respectively such that \be\label{Eq:cons-3}\bQ_{a,i}+\bQ_{b,i} = \bQ_i,\ee the charge at the boundary of the throat. Furthermore, we require that 
\be\label{Eq:cons-4}\bQ_{a,i} = \bQ_{b,{i-1}},\ee  
and thus 
$\bQ_{b,i} = \bQ_{a,{i+1}}$. Aside from these geodesics and the asymptotic boundary, the throats are bounded by light-like geodesics forming pant-like structures. We glue these $AdS_2$ pants along the geodesics as follows: we glue the edge with the $\bQ_{a,i}$ and $\bQ_{b,i}$ charges at the $i-$
 site with the edge carrying $\bQ_{b,{i-1}}$ charge in the $(i-1)$ 
  neighboring site and the edge carrying $\bQ_{a,{i+1}}$ charge in the $(i+1)$ neighboring site respectively.\footnote{We can take the asymptotic boundaries to be sufficiently large by embedding the physical patch in global $AdS_2$ and cutting out a large portion of it. This can ensure that the edges being glued have same proper lengths. This is however not necessary as will be clear from the discussion below. We prefer to keep the boundary to be the domains of the maps $u\rightarrow t_i(u)$.} This has been illustrated on a chain with $3$ lattice sites in Fig.~\ref{Fig:Network-AdS2}. We will soon discuss how this network can be the dual of as a matrix-product-state (MPS) type network of a chain of SYK-pure spin states as shown in Fig. \ref{Fig:Network-SYK}.  

The explicit details of the the trajectory of the geodetic edges specified via the respective $SL(2,R)$ charges are available in the literature (see \cite{Maldacena:2017axo} for instance). In fact, our networks also share some features with replica wormholes discussed in \cite{Penington:2019kki}.\footnote{An important difference is that we do not have the island regions which allows us to have a pant-like structure with two end-of the-world branes. The physical interpretation of our network is very different from that of replica wormholes.} We do not discuss further details here because we will relegate a complete study of the dynamics of such networks in response to external perturbations to the future. Nevertheless, it is easy to point out a trivial case where the equations of motion \eqref{Eq:Model-Shocked} remain completely unaffected in the presence of external shocks. This happens if we allow complete transmission of the shocks as they pass across the geodetic junctions from one throat to another and also perfect reflecting conditions at the boundaries of each throat. It is then possible for the shocks to traverse the full network multiple times (see Fig. \ref{Fig:Network-AdS2}) without affecting the $SL(2,R)$ charges at the glued edges or at the boundaries, except at the moment of entry into the network as in \eqref{Eq:Model-Shocked}. There is just one inconsequential complication. When the shock traverses a junction, it enters the new throat at a shifted time \cite{Shenker:2013pqa,Engelsoy:2016xyb} as shown in Fig. \ref{Fig:Network-AdS2} -- this simply follows from the Israel junction conditions. If this shift is sufficiently large, then the shock cannot exist in the network any more. In this case, for the sake of energy conservation, we can prescribe a complete absorption of the shock at the glued edge (and then $\delta \bQ$, the $SL(2,R)$ charge of the shock, should be added to the $SL(2,R)$ charge of the junction post absorption.) In any case, this also does not affect the boundary dynamics of $\bQ_i$ and thus the equations \eqref{Eq:Model-Shocked} remain completely unaffected.

For other transmission coefficients of the junctions and reflection coefficients of the boundaries, the equations \eqref{Eq:Model-Shocked} will be modified. When the shock impacts the boundaries while traversing through the network, the $SL(2,R)$ charges of the impacted boundaries are modified -- the corresponding masses $M_i$ can be either reduced or enhanced. Note the network has a self-consistent causal structure which is to be determined by the full dynamics. The full system is likely to relax to a microstate after extinction of the shocks propagating through the networks (which can be arranged by ensuring appropriate transmission and reflection coefficients mentioned). It will be challenging and intriguing to understand the dynamics of such networks. Instead of pursuing this here, let us describe analogous networks involving chains of $SYK$-spin states.

The MPS-type ansatz \cite{MPS} of SYK spin-states on a chain can be constructed as follows. At each site, we assume that there are $2N_i$ Majorana fermions $\Psi^i_k$ with $k = 1, 2, \cdots , 2N_i$ evolving via the SYK Hamiltonian. As discussed in \cite{Kourkoulou:2017zaj}, we can project any pure state of the SYK dot at each site to a state of $N_i$ spins. The Majorana fermions satisfy the anti-commutation algebra $$ \{ \Psi^i_k, \Psi^j_l\} = \delta^{ij} \delta_{kl}.$$
For any pair of Majorana fermions $\psi_1$ and $\psi_2$, we readily note that the algebra formed by $\psi_1$, $\psi_2$ and $2i\psi_1 \psi_2$ is similar to that of the Pauli matrices. We can consider the two-dimensional Hilbert space spanned by the $\vert + \rangle$ state satisfying 
$$ 2i\psi_1 \psi_2\vert + \rangle = \vert + \rangle, \quad (\psi_1 - i \psi_2)\vert + \rangle = 0,$$
and the $\vert - \rangle$ state satisfying 
$$ 2i\psi_1 \psi_2\vert - \rangle = -\vert - \rangle, \quad (\psi_1 + i \psi_2)\vert - \rangle = 0.$$
Let us define $$\mathcal{S}^i_k  = 2 i \Psi^i_{2k-1}\Psi^i_{2k}$$ with $k = 1, 2, \cdots  N_i$. We note that $$[\mathcal{S}^i_k, \mathcal{S}^j_l] = 0 \quad {\rm if} \quad i \neq j \,\,  {\rm or} \,\, i =j, \, k \neq l. $$ So we can consider simultaneous eigenstates of $\mathcal{S}^i_k$ at a given site satisfying $$ \mathcal{S}^i_k \vert B^i\rangle = s^i_k\vert B^i\rangle $$ with $ s^i_k = \pm 1.$ The $MPS$ network state can be constructed out of a chain of these SYK-spin states as illustrated in Fig. \ref{Fig:Network-SYK}. We bi-partition the $N_i$ spins at each site into $N_{a,i}$ and $N_{b,i}$ spins such that \be\label{Eq:cons-1} N_{a,i} + N_{b,i} = N_i .\ee  Also \be \label{Eq:cons-2}N_{a,i} = N_{b,{i-1}},\ee and thus $N_{b,i} = N_{a,{i+1}}$. We then maximally entangle the corresponding blocks spins with those at the nearest neighbors. 

For a concrete example, consider a chain of only 2 sites and with 2 spins (4 Majorana fermions) at each site. Our construction then results in the following (non-normalized) MPS state if we maximally entangle via forming Bell pairs:
\begin{align}
\vert B^1; B^2\rangle_1 
&= \vert ++; --\rangle 
-\vert +-; +-\rangle\nonumber\\
&-\vert -+; -+\rangle
+\vert --; ++\rangle .
\end{align}
After partitioning the two spin qubits at each site into one spin qubit each above, we have formed Bell pairs as prescribed. We note that we can maximally entangle two qubits also without forming Bell pairs:
\begin{align}
\vert B^1; B^2\rangle _2&= \vert ++; --\rangle +\vert +-; +-\rangle\nonumber\\
&+\vert -+; -+\rangle
+\vert --; ++\rangle, \nonumber\\
\vert B^1; B^2\rangle _3&= \vert ++; ++\rangle +\vert +-; -+\rangle\nonumber\\
&+\vert -+; +-\rangle
+\vert --; --\rangle, \nonumber\\
\vert B^1; B^2\rangle _4&= \vert ++; ++\rangle -\vert +-; -+\rangle\nonumber\\
&-\vert -+; +-\rangle
+\vert --; --\rangle.
\end{align}
For more sites with more spin qubits at each site, we can similarly form many possible network states with a given set of $\{N_{a,i}, N_{b,i}\}$ satisfying \eqref{Eq:cons-1} and \eqref{Eq:cons-2}. We label each such possibility by $\alpha$ and consider the following ensemble of such states:
\be\label{Eq:TN}
\rho_{\{N_{a,i}, N_{b,i}\}} = \sum_\alpha \vert c_\alpha\vert^2 e^{-\sum_i \beta_i H_i}  \vert \{B^i\}\rangle_\alpha \langle \{B^i\}\vert_\alpha e^{-\sum_i \beta_i H_i}
\ee
with $c_\alpha$ being the coefficient of normalization. Also $H_i$ is the SYK Hamiltonian at $i^{th}$ site given by
\begin{align}\label{Eq:SYK-H}
H_i &= \sum_{1 \leq a < b < c <d \leq 2 N_i} j^i_{abcd} \Psi^i_a \Psi^i_b \Psi^i_c \Psi^i_d, \\ &{\rm with} \quad \langle j^i_{abcd}\rangle^2 = \frac{3! {J_i}^2}{N_i^3}.\nonumber
\end{align}
Following \cite{Kourkoulou:2017zaj} we evolve each $\vert \{B^i\}\rangle_\alpha $ in \eqref{Eq:TN} with the local Euclidean Hamiltonian so that we can lower the energy of these states\footnote{Note that the Euclidean evolution by SYK Hamiltonian also symmetrizes over the $N_i$ spins at any given site and different ways of pairing $2N_i$ Majorana fermions at each site (with appropriate sign factors) into $N_i$ pairs each of which is projected onto a spin state. }. We consider an ensemble of such states simply because we want to consider a macroscopic state specified by the set of entanglement parameters $\{N_{a,i}, N_{b,i}\}$ and the local temperatures $\beta_i$. 

We note that the set $\{N_{a,i}, N_{b,i}\}$ is analogous to $\{\bQ_{a,i}\dt \bQ_i, \bQ_{b,i}\dt \bQ_i\}$. 
We can readily evolve the network state \eqref{Eq:TN} of the chain of SYK spin states along with the mobile $SL(2,R)$ charges $\bq_i$ using the quantum evolution equations \eqref{Eq:Model-Shocked} where we need to replace $M_i$ with $H_i$, the SYK Hamiltonian \eqref{Eq:SYK-H} at each site. These evolution equations thus take the form:\footnote{To find the $\bQ_i$ explicitly in the SYK system we need to map it to a CFT. See  eg \cite{Das:2017pif} for a discussion.}
\ba\label{Eq:SYK-Model-Shocked}
H_i' &= &- \la\(\bQ_{i-1}+ \bQ_{i+1}-2 \bQ_i\)\dt \bq_i' + \sum_A e_{i,A} \delta(u - u_{i,A}),\nonumber\\
\bq_i'' &=&\frac{1}{\si^2}\( \bq_{i-1} + \bq_{i+1} - 2 \bq_i\) \nonumber\\
&+&  \frac{1}{ \la^2}\(\bQ_{i-1}+ \bQ_{i+1}-2 \bQ_i\).
\ea
It is also obvious that the conserved energy is  
\be
\big\langle \sum_i H_i + \me_{\q}\big\rangle 
\ee in absence of perturbations and $e_{i,A}$ can be promoted to specific operators. In the future, we would like to study the dynamics of such MPS tensor network of SYK-spin states. Irrespective of whether there is a precise duality between the MPS network of SYK spin states and the $AdS_2$ throat network, the dynamics of both need to be understood individually.

 \section{Conclusions and Outlook}\label{Sec:Conc}
 
 We conclude that our simple model of a lattice of nearly $AdS_2$ throats coupled with gravitational hair in the form of $SL(2,R)$ charges 
 has desirable phenomenological features which give insights into the quantum black hole as an information processor. 
 
{Our phenomenological model reproduces the energy-absorbing and relaxation properties of a semi-classical black hole. Infalling shocks on any microstate in our model drive the transition to another microstate. Crucially, a part of the hair oscillations decouple from the black hole interior. The information in the infalling bits are encoded rapidly in these decoupled hair oscillations by the black hole interior demonstrating the phenomenon of \textit{information mirroring}. In the continuum limit, the latter absorbs almost all the energy in the infalling  bits while the energy in the hair charges is conserved on its own to a very good approximation. A part of the hair charge energy, which is stored in the form of potential energy of the configuration locked with the pre-existing black hole interior state, is unlocked by the dynamical evolution to produce the decoupled oscillations in which the information in the infallling bits is \textit{mirrored}.}
 
{We emphasize that information-mirroring is a highly non-trivial feature of our model as \textit{we can decode the classical information encoded in the impact locations and time-ordering of the infalling bits} in the decoupled hair oscillations \textit{without requiring access to any information about the black hole interior}. Although the knowledge of the pre-existing and final black hole interior microstates are not necessary, the decoding does require the knowledge of the pre-existing monopole charge of  the decoupled \textit{early} hair radiation. The phase differences of the normal modes of the final decoupled hair oscillations at each site turn out to be precisely the measurements that we need to perform (defining the code subspace) in order to get all the information necessary for the decoding.  Thus our model provides a precise realization of the Hayden-Preskill protocol.}
 
 We have discussed how our model can be generalized such that the disjointed $AdS_2$ throats can be connected via wormholes to form networks and how the latter can be interpreted as the dynamics of a tensor network of a chain of SYK-spin states. 
 
 The Page curve of our quantum black hole model can be computed in the semi-classical limit, i.e. the limit in which $N$ is large in all the $AdS_2$ throats. Particularly, it should be interesting to explicitly compute the Hawking radiation in the semi-classical limit in the networked version of our model and check if it has the features of quantum pseudorandomness as postulated by Kim, Tang and Preskill \cite{Kim:2020cds}. The latter can then ensure resolution of the AMPS type paradoxes. We also need to address some other basic questions like thermodynamic properties of our models. These, of course, require substantial work which we defer to the future. 
 
 
Our models also point in an interesting direction of \textit{fragmented holography}, a phenomenologically oriented holographic framework in which instead of considering a smooth $(d+1)-$dimensional spacetime as a dual of state in a $d-$dimensional strongly interacting quantum theory, we build a lattice of two-dimensional spacetimes which interact with each other by degrees of freedom propagating on the lattice and can be additionally networked via wormholes. It will be necessary to glue the two dimensional spacetimes on the $d-$dimensional lattice to a $(d+1)-$dimensional regular asymptotic geometry via appropriate boundary conditions to get the right features of the dual quantum theory in the ultraviolet holographically. Since the fragmentation instability of the near-extremal horizon\footnote{For a review on other kind of instabilities of non-supersymmetric horizons see \cite{Ooguri:2016pdq}.} which motivates our models is part of the usual holographic framework, fragmented holography can be relevant for the application of holography to the physics at thermal scales or at scales larger than the typical interparticle separation. Similarly, one can consider fragmented semi-holography in which the asymptotic region of the geometry beyond the fragmented horizon is replaced by weakly coupled perturbative physics, see e.g.~\cite{Faulkner:2010tq,Mukhopadhyay:2013dqa,Iancu:2014ava,Mukhopadhyay:2015smb,Banerjee:2017ozx}. 
 
 There are independent phenomenological motivations for our models particularly if the physics of the quantum system is  strongly interacting only locally. This is actually a feature of semi-local non-Fermi liquids where the equal-time correlation functions can have the usual structure as in perturbative systems but the spectral function and other unequal time correlation functions do not have quasi-particle characteristics. The instanton liquid of QCD has been also been argued to have similar features -- the instantons in a certain range of temperatures appear to be strongly correlated on the thermal circle but not along the spatial directions \cite{Schafer:1996wv}. A variation of our model may capture some features of real-time thermal QCD in a range of temperatures.\footnote{See \cite{Hellerman:2002qa} for a somewhat related discussion.} 
 
 It has been pointed out in \cite{Joshi:2019csz}  that we can map the $t-J$ model describing a strange metal to a lattice of SYK quantum dots. Furthermore, since a global $SL(2,R)$ symmetry is preserved, the time-reparametization mode gives us resistivity linear in temperature at arbitrarily low temperatures \cite{Guo:2020aog}. This crucial aspect of a global $SL(2,R)$ symmetry is also shared by our network constructions. We can add freely propagating electrons to our lattice which hybridize with a bulk fermion at each $AdS_2$ throat as discussed in \cite{Faulkner:2010tq,Mukhopadhyay:2013dqa,Doucot:2017bdm,Ben-Zion:2017tor}. In the future we would like to understand thermodynamics and electronic transport properties of such systems. 
 
 For most of the directions mentioned here, we should couple pure JT gravity to matter in each nearly $AdS_2$ throat. The study of the behavior of quantum fields in the $AdS_2$ throats networked via wormholes is a fascinating problem, relevant for a better understanding of semi-classical quantum gravity.

\acknowledgments

We are grateful to Suresh Govindarajan, Vishnu Jejjala, Lata Joshi, Arnab Kundu, Gautam Mandal, Giuseppe Policastro and Alexandre Serantes for helpful conversations; and Lata Joshi, Anton Rebhan and Pratik Roy for comments on the manuscript. TK is supported by the Prime Minister's Research Fellowship (PMRF) award of the Ministry of Human Resources and Development of India.  AM acknowledges support from the Ramanujan Fellowship and Early Career Research award of Science, Education and Research Board of the Department of Science and Technology of India; and also generous support from the New Faculty Seed Grant of Indian Institute of Technology, Madras. AS is supported by the Erwin Schr{\"o}dinger Fellowship award of the Austrian Science Fund (FWF), project no. J4406. HS acknowledges support from the INSPIRE PhD Fellowship award of Ministry of Science and Technology of India.

\appendix
\section{JT dilaton gravity and the $AdS_2$ throats}\label{App:Dilaton}
The action of 2 dimensional JT gravity is
\be
S = \frac{1}{16\pi G}\left[\int {\rm d}^2x \sqrt{-g}\Phi\left(R + \frac{2}{l^2}\right)+ S_{matter}\right].
\ee
The equations of motion of $\Phi$ and $g_{\mu\nu}$ are
\ba\label{Eq:EOM1}
R  + \frac{2}{l^2} = 0
\ea
and 
\ba\label{Eq:EOM2}
\nabla_\mu \nabla_\nu\Phi - g_{\mu\nu} \nabla^2 \Phi + \frac{1}{l^2} g_{\mu\nu} \Phi  + T_{\mu\nu}^{matter} = 0.
\ea
Note the conservation of the matter em-tensor, i.e. $\nabla^\mu T_{\mu\nu}= 0$ leads to the Bianchi identity (the above equation being divergence-free) provided $R = - 2/l^2$ 
We choose $T_{\mu\nu}$. Therefore, the above tensor equation is really one equation in disguise.

In the ingoing Eddington-Finkelstein coordinates, the metric takes the following form, setting $\kappa = 1$,
\be
{\rm d}s^2 = - 2\frac{l^2}{r^2}{\rm d}u{\rm d}r - \(\frac{l^2}{r^2}- M(u) l^2\){\rm d}u^2.
\ee
It is easy to see that the metric is locally $AdS_2$ i.e. it satisfies \eqref{Eq:EOM1}.

For application to the shock wave geometry we choose $T_{\mu\nu}$ of the form:
\be\label{Eq:Tmn-throat}
T_{uu} = f(u), \quad T_{ur} = T_{ru} = T_{rr} = 0.
\ee
It is easy to check that $\nabla^\mu T_{\mu\nu}= 0$ identically.

The $rr-$component of \eqref{Eq:EOM2} then yields:
\be
\partial_r^2 \Phi + \frac{2}{r}\partial_r\Phi = 0.
\ee
The solution to the above is
\be\label{Eq:Phisol1}
\Phi = \frac{a(u)}{r}+ b(u).
\ee
With the above form of $\Phi$, the $ru-$component of \eqref{Eq:EOM2} yields
\be\label{Eq:Phisol2}
b(u) = a'(u).
\ee
Substituting both \eqref{Eq:Phisol1} and \eqref{Eq:Phisol1} in the $uu-$component of \eqref{Eq:EOM2} yields 
\be
a'''(u) -  a'(u) M(u)- \frac{1}{2}a(u)M'(u) + f(u) = 0.
\ee
We set the Dirichlet boundary condition: $a(u) = \phi_r$ is a constant. We also realize that setting $\kappa = 1$ amounts to setting $\phi_r = 2$ so that $a(u) = \phi_r$. This implies that
\be
M'(u) = f(u).
\ee

This equation holds for each $AdS_2$ throat. In order to match with \eqref{Eq:Model-Shocked}, we need to set
\be
M_i'(u) = f_i(u).
\ee
with
\begin{align}
f_i(u) = &- \la\(\bQ_{i-1}(u)+ \bQ_{i+1}(u)-2 \bQ_i(u)\)\dt \bq_i'(u) \nonumber\\ &+ \sum_A e_{i,A} \delta(u - u_{i,A}).
\end{align}
The inter-throat coupling thus gives rise to a time-dependent infalling or outgoing dilute null matter in each throat along with the ingoing shocks. As clear from \eqref{Eq:Tmn-throat} we must have 
\be\label{Eq:Tmn-throat-2}
T_{(i)uu} = f_i(u), \quad T_{(i)ur} = T_{(i)ru} = T_{(i)rr} = 0.
\ee
The above em-tensor is conserved in each throat metric:
\be
{\rm d}s_{i}^2 = - 2\frac{l^2}{r^2}{\rm d}u{\rm d}r - \(\frac{l^2}{r^2}- M_i(u) l^2\){\rm d}u^2.
\ee
Remarkably, the dilaton in each throat takes the simple form 
\be
\Phi_{(i)}(r,u)  = \frac{2}{r}.
\ee
In this discussion, we have skipped the details of holographic interpretation of JT-gravity. We refer the reader to \cite{Almheiri:2014cka,Jensen:2016pah,Engelsoy:2016xyb,Maldacena:2016upp} for these details and also to \cite{Cvetic:2016eiv,Grumiller:2017qao} for a broader discussion on two-dimensional quantum gravities with a dilaton.

\section{Drift and phase shift}\label{App:Drift}
Below we present some numerical details pertaining to the removal of numerical noise essential for extracting the phases in the Fourier transforms discussed in Sec. \ref{Sec:HP}. We will first discuss the numerical drift that is present in the data for $q_{i}'$, its origin and how we subtract it. We then present some details about how we compute the phase differences that are used in the decoding procedure and discuss why computing these accurately becomes harder for six or more sites due to numerical noise.

\subsection{Correcting the drifts}

\begin{figure}\centering
\includegraphics[width=\linewidth]{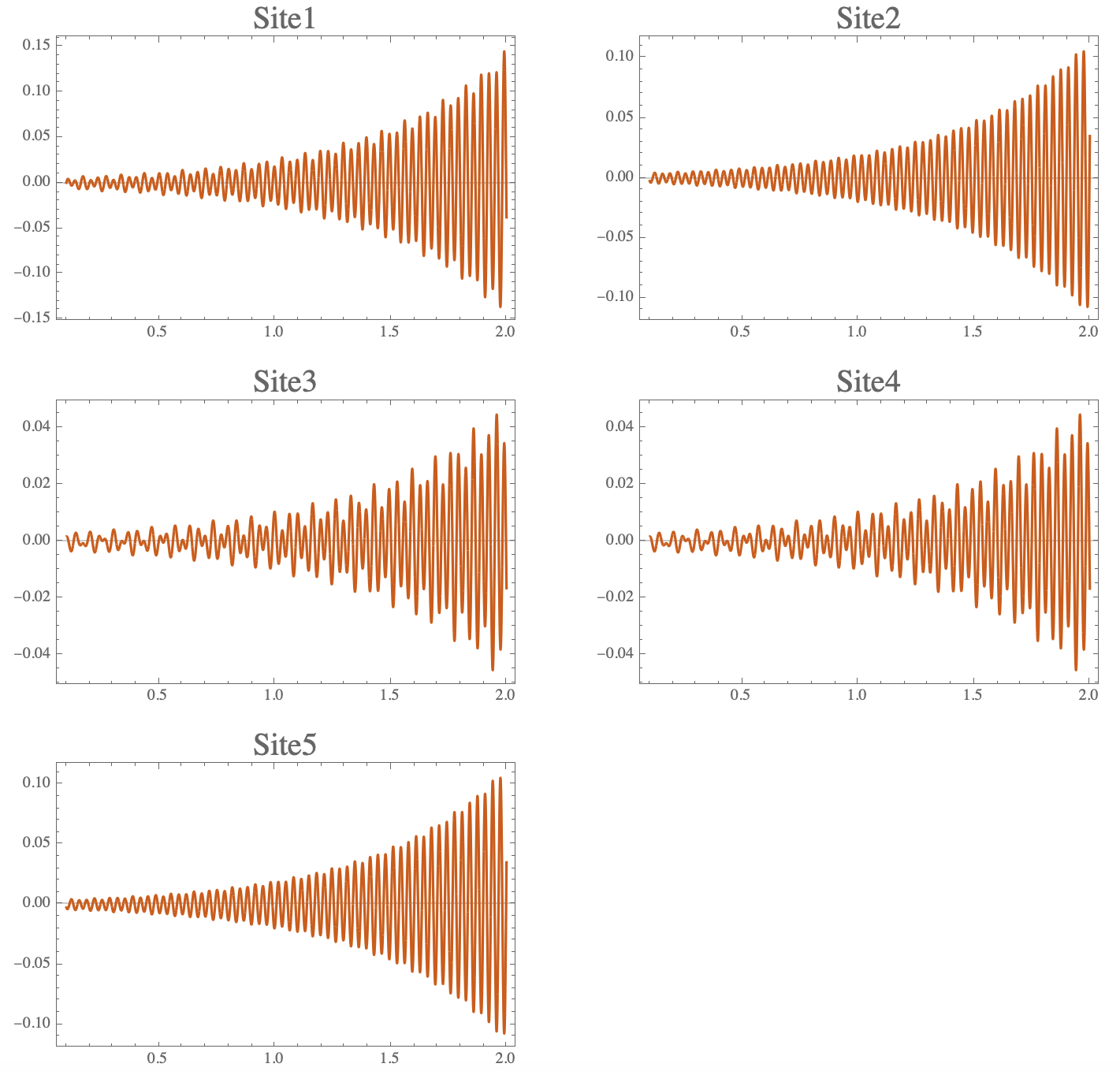}\caption{Plot of the $q_{i}'$ after a single shock at site 1. The amplitude of the oscillations can be clearly seen to grow with time. }\label{Fig:Drift1}
\end{figure}

The decoding procedure involves studying the $q_{i}'$ carefully, computing their Fourier transform and then extracting the phase differences between the two frequency modes present at each site. The $q_{i}'$ are shown in Fig. \ref{Fig:Drift1}.

\begin{figure}
\begin{minipage}{\linewidth}\centering
\includegraphics[width=\linewidth]{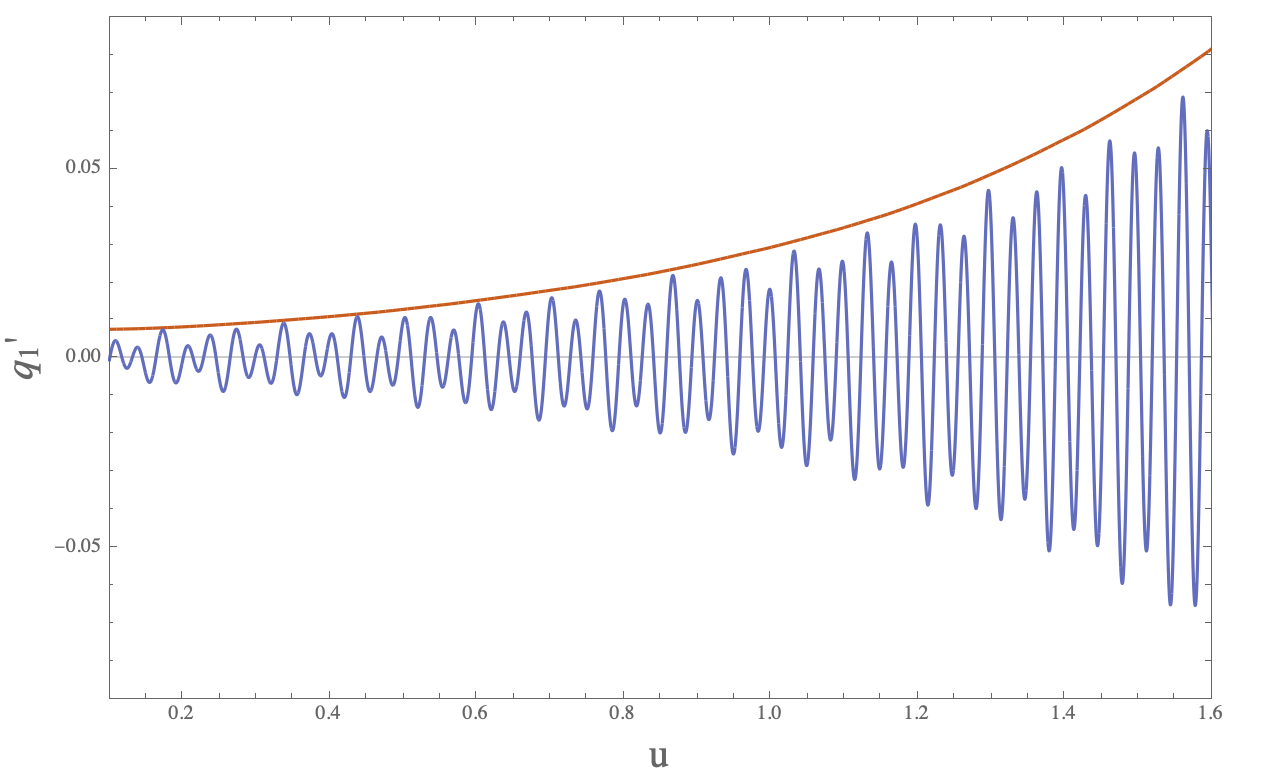}\caption{The envelope of $q_{i}'$ is fit to an exponential (red curve) }\label{Fig:Drift2}
\end{minipage}\;
\begin{minipage}{\linewidth}\centering
\includegraphics[width=\linewidth]{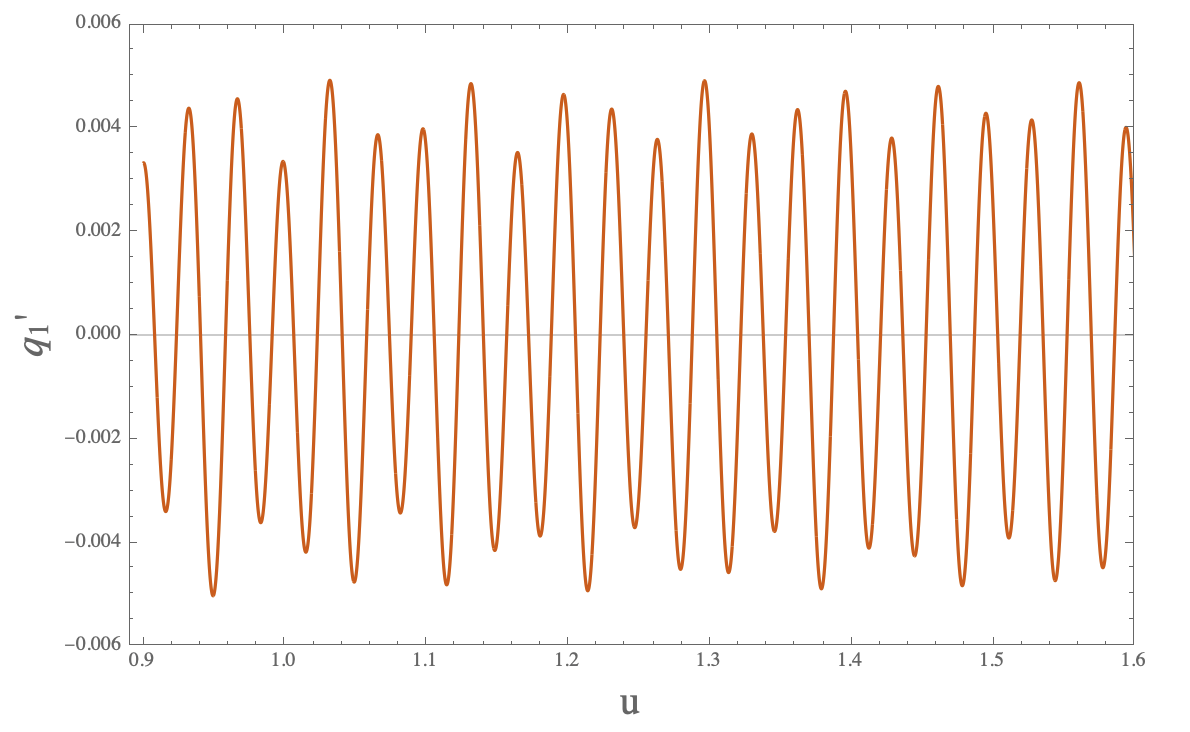}\caption{$q_{i}'$ after the drift is subtracted out.}\label{Fig:Drift3}
\end{minipage}
\end{figure}

The $q_{i}'$ are oscillating functions with an amplitude that grows in time. This growth in amplitude is a consequence of the fact that all the $\Q^{(0)}$ in the right hand side of \eqref{Eq:Model} are not exactly equal due to numerical noise. Thus the right hand side of the equation for $\q^{(0)}$ has an extra term because of numerical error. This extra term results in the growth of the amplitude of $q_{i}'$ seen in Fig.~\ref{Fig:Drift1} via resonant feedback. We can remove this growth of amplitude by fitting the envelope of the $q_{i}'$ functions to an exponential. We then multiply $q_{i}'$ with the inverse of this exponential to eliminate this growth. This process is illustrated in Fig.~\ref{Fig:Drift2} for one site in the model. The data for $q_{i}'$ after correcting for this drift is plotted in Fig.~\ref{Fig:Drift3}.

\subsection{Compensating for the jumps}
\begin{figure}
\centering
\includegraphics[scale=0.4]{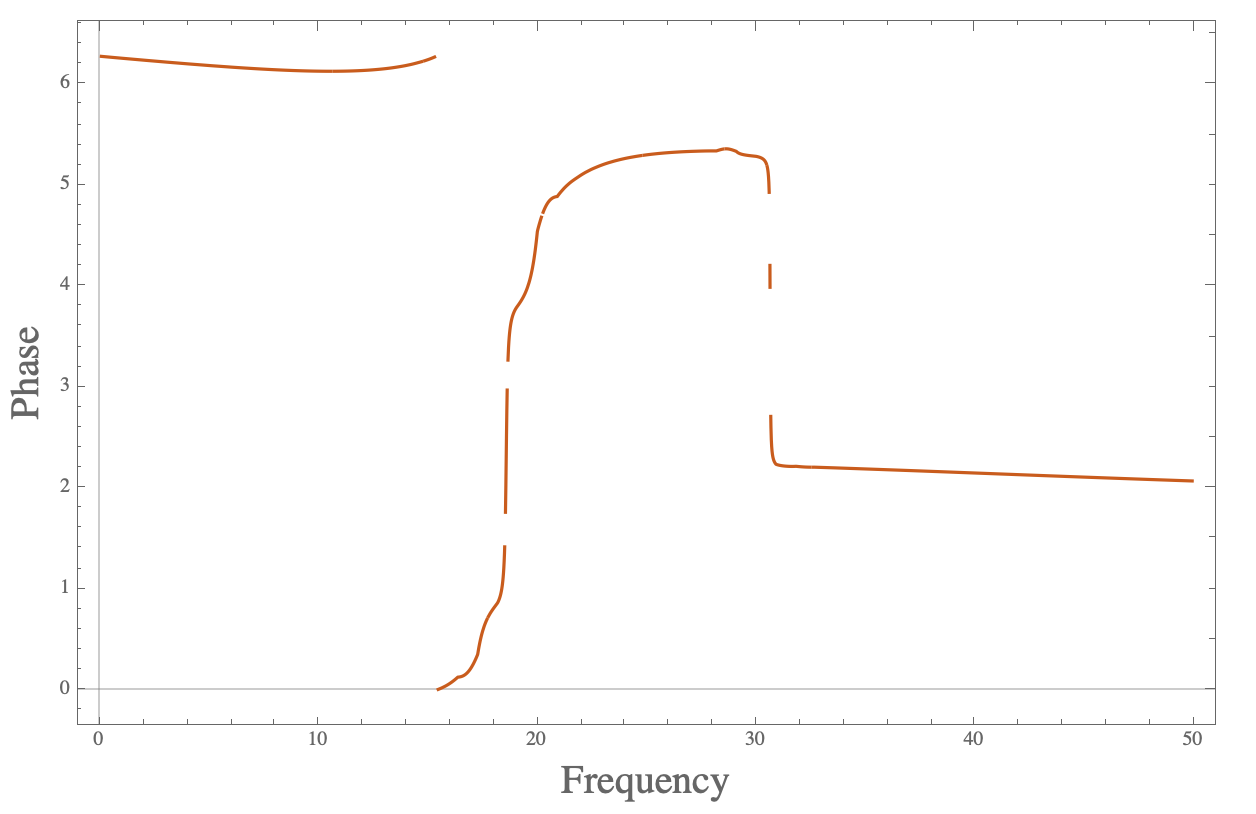}\caption{Plot of the phase against the frequency for site 1. The gaps in the plot are due to interpolation. Note that the phase is computed modulo $2\pi$}\label{Fig:Phase}
\end{figure}

After correcting for the drift in amplitude of $q_{i}'$, the next step in our decoding procedure is to compute the Fourier transform of $q_{i}'$ to extract the phase differences between the two frequency modes present at each site. The interpolated data for the phase vs frequency is shown for site 1 in Fig. \ref{Fig:Phase}.

Recall that the $q_{i}'$ obey the discrete Klein Gordon equation given in \eqref{Eq:KGj} and the normal mode frequencies for this equation if $\si = 0.01$ are $30.2731$ and $18.7098$.  In Fig. \ref{Fig:Phase} one can clearly see that the phase jumps by approximately $\pi$ at the values of the normal mode frequencies. This can be explained by recalling that 
There is a small forcing term on the right hand side of \eqref{Eq:KGj} due to numerical noise. This in turn should cause a jump in the phases by $\pi$ at the normal mode frequencies due to resonant feedback. The jumps are sharp due to the absence of damping. When computing the phases for the decoding protocol, we need to be careful about these $\pi-$discontinuities in the phases. One simple way to compensate for these discontinuities is to measure the phases on the correct side of the discontinuities which are exhibited in Fig.~\ref{Fig:Phase}. The phase difference should be computed between a point slightly to the left of the jump at the first mode and a point slightly to the right of the jump at the second mode ensuring that the $\pi-$discontinuities have no effect on the phase differences we are computing.

As we increase the number of sites, we expect the number of normal modes to grow. For instance, in a chain with 6 sites we will have 3 distinct normal mode frequencies. For $\si = \frac{6}{5} \times 0.01$, the normal mode frequencies are: $(26.526, 22.972, 13.263 )$. In this case the frequencies are much closer together compared to the 5 site case. Since the highest normal mode frequencies are close to each other, the numerical noise 
can cause the $\pi-$discontinuities to overlap. It is then hard to compute the phases of these normal modes accurately, which meant that we were unable to check if our decoding generalizes to higher number of sites. 

\section{Decoding for 3 and 4 site models}\label{App:Decode34}
Here we briefly discuss the decoding procedure for a chain model with lower number of sites than the 5 site model discussed in Sec.~\ref{Sec:HP}.
For a chain with 3 sites we expect only one distinct normal mode frequency for the discrete Klein Gordon equation \eqref{Eq:KGj}. We can decode the position and time ordering of the shocks by studying the phase of this frequency mode for each site. In the case of a single shock to a single site we observe the same symmetry pattern we have seen before for the 5 site case. If Alice shocks two distinct sites in a 3 site chain, the positions of the shocks are given by the positions of the extrema of the phases. The site with the minimum phase was shocked first and the site with the maximum phase was shocked second. 

The decoding for a chain with 4 sites is slightly different from the 3 and 5 site cases. We have two distinct normal mode frequencies just like the 5 site model. For a single shock to a 4 site chain we again observe the same symmetry pattern we have seen before. For two shocks to antipodal lattice points (1,4 or 2,3) we observe the same symmetry pattern as a single shock. For instance the pattern for a single shock to site 1 is the same as the pattern for shocks to sites 1 followed by 4. We do not know how to distinguish between these two cases. To decode two shocks to two nearest neighbour sites in a 4 site chain we need to study the ordering of the phase differences. The information about the position and time ordering is not encoded in the extrema of the phase differences. This information is instead encoded in the ordering of the phase differences. When phases are listed in ascending order, there is a one to one map from each such ordering to the position and time ordering of two shocks. For instance if site 1 is shocked first followed by site 2 then the phases are ordered as: $\phi_{4} < \phi_{3} < \phi_{2} < \phi_{1}$. If the time ordering of the shocks is reversed then the phases are ordered as: $\phi_{4} < \phi_{1} < \phi_{2} < \phi_{3}$. For 2 shocks, these orderings always form a closed loop in the 4 site chain. Note that the number of possible loops is 8 (two possible directions and 4 possible starting points) and the number of possible nearest neighbour shocks is also 8.

\bibliography{NAdS2LN}

\end{document}